\documentclass[a4paper]{article}

\usepackage[contents=2021\,Canadian\,Journal\,of\,Physics\,99\,222;\,doi:10.1139/cjp-2020-0313, pages=all, color=black, position={current page.south}, placement=bottom, scale=1, opacity=1, vshift=5mm]{background}

\usepackage[margin=1in]{geometry} 

\usepackage{newtxtext}
\usepackage{amsmath}
\usepackage{amsthm}
\usepackage{amssymb}

\usepackage[T1]{fontenc}

\usepackage[utf8]{inputenc}
\usepackage{hyperref}

\usepackage[sort&compress,numbers,square]{natbib}
\usepackage{graphicx, color}
\graphicspath{{fig/}}

\usepackage{mathrsfs} 

\title{The dynamics of spatially confined oscillations}
\author{Till Stadtler$^1$ \and Pavel Kroupa$^{1,2}$ \and Manfred Schmid$^3$}

\date{
	$^1$Helmholtz-Institut f\"ur Strahlen und Kernphysik, Universit\"at Bonn, Nussallee 14-16, 53115 Bonn, Germany
	\\ \texttt{pkroupa@uni-bonn.de}\\%
	$^2$Charles University in Prague, Faculty of Mathematics and Physics, Astronomical Institute, V~Hole\v{s}ovi\v{c}k\'ach 2, CZ-18000 Praha, Czech Republic\\%
	$^3$Eboracum GmbH, Im Vogelsang 9, 53343 Wachtberg, Germany
	\\[2ex]%
}

\begin{document}
	\maketitle
	
	\begin{abstract}
The possible relation of the wave nature of particles to gravitation as an emergent phenomenon is addressed. Hypothetical particles are considered as spatially confined oscillations (SCOs) and are constructed
  through the superposition of plane waves. The effect of a
  continuously changing refractive index (speed of propagation field) on SCOs is calculated and
  the continuous Ibn-Sahl--Snell law of refraction is derived. Refracted
  plane wave constituents of SCOs in an inhomogeneous medium cause the
  oscillation as a whole to accelerate as an entity. This acceleration
  is described by a geodesic equation, in much the same way as in
  general relativity. The proper time of an SCO can be defined via its
  oscillation frequency. The proper time and its change along the
  trajectory are equivalent to a particle in general relativity as
  described by the 0th component of its geodesic equation.  An SCO in
  an inhomogeneous refractive index field exhibits general relativistic properties
  based on basic wave mechanics.  Properties
  derived from direct calculations are length contraction,
  gravitational red- and blueshift and Thomas precession. The
  presented theory is an approximation for oscillations which are
  small compared to changes in the refractive index field. SCOs in an inhomogeneous medium may thus
  yield a naturally emerging particle-field interaction with general
  relativistic properties and may allow a useful vantage point on the nature of 
  gravitation using classical-wave experiments.
		
		\noindent\textbf{Keywords:} history and philosophy of astronomy -- miscellaneous -- gravitation -- elementary particles -- waves -- methods: analytical
	\end{abstract}


\section{INTRODUCTION}
\label{sec:intro}
Using information entirely based on the Solar System obtained by Brahe and Kepler and on Earth experiments by Galilei, in 1687, Newton\cite{Newton1687} suggested an empirically-derived classical law of universal gravitation which is today interpreted to be an effective mathematical formulation allowing the motions of astronomical mass entitites to be accurately and precisely calculated and predicted. It is based on the classical notion of instantaneous action at a distance.  The modern nature of gravitation is generally understood according to Einstein's \cite{Einstein16} theory of general relativity to result from matter curving space-time. This description has been shown to be extraordinarily successful in the strong-field regime (Solar System, neutron stars, black holes). When applied to the spatial scales of galaxies and larger, a missing mass problem appears, needing the introduction of dark matter and dark energy which together dominate the energy content of the Universe in the current Standard Model of Cosmology. These constituents are, however, not described by the Standard Model of Particle Physics. In 2017, Verlinde \cite{Verlinde17} notes this as a motivation to reconsider the fundamental theory underlying the phenomenon gravitation. Prior to that, in 2011 Verlinde \cite{Verlinde11} suggested that space is an emergent quantity in a 
holographic scenario and that gravitation can be viewed as an entropic force which results from changes in the information linked to the positions of particles. Given the hitherto not well understood dark matter and dark energy, Verlinde \cite{Verlinde17} studies
the emergent laws of gravitation by taking into account the volume law contribution to the entropy. By only invoking the natural constants of nature, Verlinde finds
the law of gravity to begin to deviate from Einstein's law when the astronomical observations indicate the appearance of dark matter and dark energy. Verlinde \cite{Verlinde17} thus argues to have found a possible explanation for the empirically-discovered Milgrom's critical acceleration scale $a_0$, which emerges from the new Milgromian empirical classical description of gravitation which extends Newton's efforts by including astronomical data encompassing the Solar System and galaxies \citep{Milgrom83, BM84}. 

With this study we approach the problem of gravitation from a different vantage point which, however, should neither be seen as being in contradiction to Einstein's nor Verlinde's, because it can be shown to ultimately lead to an interpretation of space-time being distorted and is as well related to information content through the oscillations.
Because particles have the property of being describable as de Broglie waves (e.g. \cite{BarOr+19}), we take a step back and consider a highly simplified and hypothetical, idealised (toy) universe, in which plane waves can propagate as disturbances. As a simplification and for the sole purpose of computational ease, we assume the waves propagate in an ideal classical physical medium which has a propagation speed, $c_s$, for these waves. It is noted that it may be possible to set up classical wave experiments to test the theory developed in this contribution, and to make general relativistic effects accessible to classical laboratory experiments.

An oscillation is a strictly causal and logical phenomenon allowing energy to propagate. The superposition of such waves can produce interference patterns which can exhibit interesting properties that may be educational, if not relevant for the quantum-mechanical concept of the particle-wave duality.  An explorative step in this direction was taken in \cite{schmidkroupa2014}, who constructed a standing spherical wave (referred to as a {\it spheron}) from a superposition of incoming and outgoing spherical waves. For a moving spheron to remain a solution of the wave equation it needs to be actively Lorentz transformed. Spherons are localised, carry energy, are naturally restricted to propagate at a velocity smaller than $c_s$ and quantum mechanical operators quantify the position and momentum of spherons.  A {\it spheronic toy universe} can thus be constructed, being composed of ``particles'' which are spherons. In the spheronic toy universe, it follows that if all matter consists of these oscillations, including observers and their measurement tools, then this would explain the null result of a hypothetical Michelson-Morley experiment \citep{MM} if it were performed in the spheronic toy universe: the wave-carrying medium cannot be inferred on the basis of this experiment.  The natural emergence of the Lorentz transformation in this context makes it worthwhile to dwell deeper into this seemingly simple concept.

In this contribution, the spheron is extended to an arbitrary spatially confined oscillation (SCO). All properties found in \cite{schmidkroupa2014} for spherons apply to SCOs. These include the idea that localised oscillations of the medium represent particles (in the spheronic toy universe) and the concept of an observer with proper time and features similarly found in quantum mechanics.  To extend the spheronic toy universe to include the dynamics of particles, we derive a mechanism to accelerate SCOs by introducing an inhomogeneous speed of propagation field.  The term ``speed of propagation field'' is used, because the physical principle, on which this contribution is based, is classical wave mechanics which eases the computations.  The concept of an observer with proper time is applied to SCOs in homogeneous and inhomogeneous speed of propagation fields.  The notion followed here is to see how (and if) this may allow to describe physical phenomena, such as gravity or any particle-field interaction. If this is possible, then the ansatz followed here may allow an additional vantage point towards interpreting gravitation. 

The concept of decomposing periodic oscillations into plane waves is explained in detail in Sec.~\ref{sec:basicknowledge}, among other concepts, such as the Lorentz transformation and Thomas-Wigner rotation. The properties of plane waves, as constituents of SCOs, in an inhomogeneous speed of propagation field are derived in Sec.~\ref{sec:methods}, which results in a continuous Ibn-Sahl--Snell's law of refraction. This law describes a continuous Lorentz transformation of plane waves in inhomogeneous mediums. Sec.~\ref{sec:results} derives the complete dynamics of SCOs in arbitrary inhomogeneous speed of propagation fields in the form of an equation of motion of an SCO. This equation of motion and the change of the proper time of an SCO are equivalent to the geodesic equation in general relativity, see Sec.~\ref{subsec:geodesictrajectory}. A conclusion is given in Sec.~\ref{sec:conclusion}.

\section{Some background}
\label{sec:basicknowledge}

Plane wave decomposition of an arbitrary oscillation is introduced in Sec.\ \ref{subsec:planewave}. It is used throughout this contribution to describe how an arbitrary oscillation is affected by a change of its plane wave constituents. In Sec.\ \ref{subsec:lorentz}, Lorentz transformations in arbitrary directions are explained. They are used later to actively transform oscillations. Two subsequent non-collinear Lorentz transformations are described by Thomas-Wigner rotation, presented in Sec.\ \ref{subsec:wigner}. This work is briefly placed into historical context in Sec.~\ref{subsec:history}.

\subsection{Plane Wave Decomposition}
\label{subsec:planewave}
The plane wave decomposition is similar to the usage of the Fourier transform in periodic signal processing \citep{bracewell2000fourier}. To apply it to an oscillation in space, the concept is extended to three dimensions.

Any arbitrary periodic oscillation in a homogeneous, isotropic medium with speed of propagation $c_s$ which solves the classical wave equation,
\begin{equation}
\frac{1}{c_s^2}\partial^2_t\rho(\mathbf{x},t) - \Delta\rho(\mathbf{x},t) = 0,
\label{eq:waveeq}
\end{equation}
can be decomposed via Fourier transform into a superposition of plane waves. The density fluctuation at a three-dimensional position vector $\mathbf{x}$ at time $t$ is given by \begin{equation}\label{eq:superposition} \rho(\mathbf{x},t) = \operatorname{Re}\left[\int\!\mathrm{d}^3k \,\tilde{\rho}(\mathbf{k})\exp[\mathrm{i}\mathbf{k}\cdot\mathbf{x}-\mathrm{i}\omega t]\right], \end{equation} with the complex frequency spectrum $\tilde{\rho}(\mathbf{k})$. The complex frequency spectrum contains information about the amplitude $|\tilde{\rho}(\mathbf{k})|$ and phase offset $\mathrm{Arg}[\tilde{\rho}(\mathbf{k})]$ of the plane waves. All plane waves have a constant phase velocity $v_\mathrm{p}=c_s=\omega/|\mathbf{k}|$. The integral over $\mathbf{k}$-space produces a continuous distribution of plane waves with different wave vectors $\mathbf{k}$, which contain information about the direction of the plane wave and its wavelength via $\lambda = 2\pi/|\mathbf{k}|$. An integral over frequencies is not necessary, as all possible frequencies are covered by the $\mathbf{k}$-space integral via $\omega=c_s|\mathbf{k}|$. The real part is taken to obtain real values to represent the density fluctuation. Any oscillation that can be represented by the right hand side of Eq.\ \ref{eq:superposition} solves Eq.~\ref{eq:waveeq}.

The plane wave decomposition explained here is used throughout this contribution. It is applied to describe the change of an oscillation as a whole when a change in its plane wave constituents occurs.

\subsection{Lorentz Transformation in an Arbitrary Direction}
\label{subsec:lorentz}

The Lorentz transformation is a set of equations representing a linear four-dimensional coordinate transformation, relating space and time coordinates. In the context of this contribution, the Lorentz transformation is not used to relate reference frames. Instead, it is used as an active transformation between scalar functions representing oscillations. In Sec.\ \ref{subsec:activelorentz}, we are reminded that any oscillation actively transformed via the Lorentz transformation also solves the classical wave equation. 

The Lorentz transformed space and time coordinates are linearly related to the old space and time coordinates. The coefficients relating space and time coordinates reveal a symmetry, resulting in the complete description of any Lorentz transformation. This can already be seen in the commonly used set of equations,
\begin{align}\label{Lorentz1}
z' ={}& \gamma[z-\beta c_st], \\
c_st' ={}& \gamma[c_st-\beta z],\label{Lorentz12}
\end{align}
with $\beta$ and $\gamma$ related by
\begin{equation}\label{betagamma}
\gamma = \frac{1}{\sqrt{1-\beta^2}} \Rightarrow \gamma^2 = 1 + \gamma^2\beta^2.
\end{equation}

The parameters $\beta$ and $\gamma$ represent one degree of freedom with either $\beta\in[0,1)$ or $\gamma\in[1,\infty)$. This commonly used Lorentz transformation is performed in the $z$-direction, fixing two degrees of freedom. It can be extended to three spatial dimensions, introducing the direction of the Lorentz transformation denoted as the unit-vector $\mathbf{n}$,
\begin{align}\label{Lorentz3}
\mathbf{x}' ={}& \gamma[\mathbf{x}\cdot\mathbf{n} - \beta c_st]\mathbf{n} + \mathbf{x} - \mathbf{n}(\mathbf{x}\cdot\mathbf{n}) \\
& \Rightarrow \begin{cases}
\mathbf{x}' \times \mathbf{n} = \mathbf{x} \times \mathbf{n},\\
\mathbf{x}' \cdot \mathbf{n} = \gamma[\mathbf{x}\cdot\mathbf{n} - \beta c_st],
\end{cases} \\
c_st' ={}& \gamma[c_st - \beta\mathbf{x}\cdot\mathbf{n}].\label{Lorentz32}
\end{align}

Eq.\ \ref{Lorentz3} and Eq.\ \ref{Lorentz32} reduce to Eq.\ \ref{Lorentz1} and Eq.\ \ref{Lorentz12} for $\mathbf{n}=\mathbf{e}_z$. The space coordinates perpendicular to $\mathbf{n}$ are not affected. The symmetry can be seen by writing the linear relation between the new and old space and time coordinates as a matrix:
\begin{align}\label{Lor}
\begin{pmatrix}
c_st' \\ \mathbf{x}'
\end{pmatrix} &=
\begin{pmatrix}
\gamma & -\beta\gamma\mathbf{n}^T \\
-\beta\gamma\mathbf{n} & \mathbf{I} + (\gamma-1)\mathbf{n}\mathbf{n}^T
\end{pmatrix}
\begin{pmatrix}
c_st \\ \mathbf{x}
\end{pmatrix} \\
&\equiv  \mathbf{B}(\gamma,\mathbf{n})\begin{pmatrix}
c_st \\ \mathbf{x}
\end{pmatrix},
\end{align}
with the Lorentz matrix $\mathbf{B}(\gamma,\mathbf{n})$, introduced here. This short notation is used for readability purposes.

The inverse Lorentz transformation is given by the inverse of the matrix in Eq.\ \ref{Lor}, which is equivalent to substituting $\mathbf{n} \to-\mathbf{n}$. The notations in Eq.\ \ref{Lorentz3} and Eq.\ \ref{Lor} will be used throughout this contribution, as seen fit.

\subsection{Thomas-Wigner Rotation}
\label{subsec:wigner}

The Thomas-Wigner rotation is the mathematical concept of describing two subsequent non-collinear Lorentz transformations as the combination of a single Lorentz transformation and a spatial rotation. This concept was discovered by Thomas in Ref.~\cite{Thomas:1926dy} and derived by Wigner in Ref.~\cite{Wigner1939}. The order of the single Lorentz transformation and the spatial rotation is not fixed. Thomas-Wigner rotation can be written as a Lorentz transformation first, spatial rotation second, or vice-versa. The Thomas-Wigner rotation arises in special relativity, but in the context of this contribution, it will be used for mathematical purposes only; to describe the effect of two subsequent active transformations of an oscillation.

Two subsequent non-collinear Lorentz transformations are described by six free parameters, three from each Lorentz transformation. The six free parameters have to be represented by the combination of a single Lorentz transformation and the spatial rotation; three are represented by the single Lorentz transformation, another three by the rotation axis (two parameters) and the rotation angle (one parameter) of the spatial rotation. As the concept of Thomas-Wigner rotation is already known and sufficiently proven, it will not be derived here. For calculations, see \cite{WigCalculations}. Only the relation between the six free parameters is of interest.

Each Lorentz transformation is described by three degrees of freedom, namely its transformation parameter $\beta$ or $\gamma$ and the direction. For two non-collinear transformations, the directions are different with arbitrary transformation parameters. Here, the first Lorentz transformation is described by the parameters $(\gamma_a,\mathbf{n})$, the second by the parameters $(\gamma_b,\mathbf{m})$, see Eq.\ \ref{twolorentz}. The resulting single Lorentz transformation is described by the parameters $(\gamma, \mathbf{o})$ or $(\gamma,\mathbf{p})$, depending on the order of transformation and spatial rotation, see Eq.\ \ref{Wigtwocases} and Eq.\ \ref{Wigtwocases2}. The spatial rotation matrix $\mathbf{R}(\vec{\alpha})$ is the same in both cases and is described, using axis-angle representation, by the vector $\vec{\alpha}$.\footnote{Throughout this text vectors are written in boldface except for Greek letters, the vectorial status of which is designated by an arrow.} The direction of $\vec{\alpha}$ is the rotation axis, the magnitude $|\vec{\alpha}|$ is the rotation angle. 

Two subsequent Lorentz transformations still lead to a linear relationship between old and new space and time coordinates. The new space and time coordinates are denoted as $(ct'',\mathbf{x}'')$,
\begin{equation}\label{twolorentz}
\begin{pmatrix}
c_st'' \\\mathbf{x}''
\end{pmatrix}
=
\mathbf{B}(\gamma_b,\mathbf{m})\mathbf{B}(\gamma_a,\mathbf{n})
\begin{pmatrix}
c_st \\ \mathbf{x}
\end{pmatrix},
\end{equation}
with the Lorentz matrix $\mathbf{B}$ as defined in Eq.\ \ref{Lor}

The Thomas-Wigner rotation is given by the combination of a single Lorentz transformation and a spatial rotation that results in the same transformation as the two non-collinear Lorentz transformations in Eq.\ \ref{twolorentz}: 
\begin{align}\label{Wigtwocases}
\begin{pmatrix}
c_st'' \\\mathbf{x}''
\end{pmatrix}
&\stackrel{(1.)}{=} \mathbf{B}(\gamma,\mathbf{o})
\begin{pmatrix}
1 & 0 \\
0 & \mathbf{R}(\vec{\alpha})
\end{pmatrix} 
\begin{pmatrix}
c_st \\ \mathbf{x}
\end{pmatrix},\\
\begin{pmatrix}
c_st'' \\\mathbf{x}''
\end{pmatrix}
&\stackrel{(2.)}{=} \begin{pmatrix}
1 & 0 \\
0 & \mathbf{R}(\vec{\alpha})
\end{pmatrix}
\mathbf{B}(\gamma,\mathbf{p})
\begin{pmatrix}
c_st \\ \mathbf{x}
\end{pmatrix}.\label{Wigtwocases2}
\end{align}

Interestingly, the two possible directional unit-vectors $\mathbf{o}$ and $\mathbf{p}$ are related by the same rotation matrix used in the spatial rotation. For detailed calculations see \cite{WigCalculations}.
\begin{equation}
\mathbf{R}(\vec{\alpha})\mathbf{p} = \mathbf{o} \Leftrightarrow \mathbf{p} = \mathbf{o}^T\mathbf{R}(\vec{\alpha}).
\end{equation}

This allows the spatial rotation to be described with the unit-vectors $\mathbf{o}$ and $\mathbf{p}$. Performing the matrix multiplication in both cases leads to the same matrix describing the effect of the Thomas-Wigner rotation in a single matrix as
\begin{equation}\label{Wig}
\begin{pmatrix}
c_st'' \\\mathbf{x}''
\end{pmatrix}=
\begin{pmatrix}
\gamma & -\beta\gamma\mathbf{p}^T \\
-\beta\gamma\mathbf{o} & \mathbf{R}(\vec{\alpha}) + (\gamma-1)\mathbf{o}\mathbf{p}^T
\end{pmatrix}
\begin{pmatrix}
c_st \\\mathbf{x}
\end{pmatrix}.
\end{equation}

The symmetry is broken. The upper right term and lower left term are both proportional to $\beta\gamma$ but act in different directions. The interpretation of this matrix is difficult. It is easier to think in terms of a spatial rotation and a single Lorentz transformation. 

The six degrees of freedom represented by the parameters $\beta_a$, $\beta_b$, $\mathbf{n}$ and $\mathbf{m}$ are related in a specific way. The following relations between the parameters can be verified by comparing Eq.\ \ref{Wigtwocases} and Eq.\ \ref{Wigtwocases2} with Eq.\ \ref{Wig}. The parameters are related by
\begin{align}\label{WigSix}
\gamma ={}&\gamma_a\gamma_b[1 + \beta_a\beta_b(\mathbf{m}\cdot\mathbf{n})], \\
\begin{split}
\mathbf{o} ={}& [\beta_a\gamma_a\mathbf{n}+[\beta_a\gamma_a(\gamma_b-1)(\mathbf{n}\cdot\mathbf{m}) \\
& +\gamma_a\beta_b\gamma_b]\mathbf{m}][\beta\gamma]^{-1}, 
\end{split}\\
\begin{split}
\mathbf{p} ={}& [\beta_b\gamma_b\mathbf{m}+[\beta_b\gamma_b(\gamma_a-1)(\mathbf{m}\cdot\mathbf{n}) \\
& +\gamma_b\beta_a\gamma_a]\mathbf{n}][\beta\gamma]^{-1}.\label{WigSix2}
\end{split}
\end{align}

And the rotation matrix is fixed by
\begin{align}\label{rotationparams}
\vec{\alpha}/|\vec{\alpha}| ={}& \frac{\mathbf{p}\times\mathbf{o}}{|\mathbf{p}\times\mathbf{o}|} = \frac{\mathbf{n}\times\mathbf{m}}{|\mathbf{n}\times\mathbf{m}|}, \\
\cos(|\vec{\alpha}|) ={}& \mathbf{p}\cdot\mathbf{o}. \label{rotationparams2}
\end{align}

The combined $\gamma$ describes the velocity addition of two subsequent Lorentz transformations. The resulting $\gamma$ is still found in $[1,\infty)$, which leads to a combined $\beta\in[0,1)$. The unit-vectors $\mathbf{o}$ and $\mathbf{p}$ are linear combinations of $\mathbf{n}$ and $\mathbf{m}$. Therefore, the rotation axis can be equivalently described by $\mathbf{p}$ and $\mathbf{o}$, and $\mathbf{n}$ and $\mathbf{m}$. The effect of the Thomas-Wigner rotation is discussed in more detail in Sec.\ \ref{subsec:oscinmotion}.

\subsection{A Few Notes on the Historical Context}
\label{subsec:history}

The concepts of Lorentz contraction and time dilation were developed
well before the introduction of special relativity by Albert
Einstein. Lorentz transformations were already known and heavily
discussed at the end of the 19th century due to their property of
leaving Maxwell’s equations invariant \citep{Lorentz1937}. They were
used by Hendrik Lorentz and Henri Poincaré to derive a theory based on
the concept of the luminiferous ether, which is mathematically
equivalent to special relativity \citep{Lorentz1937}. This theory is
called Lorentz ether theory (LET). In LET, Lorentz contraction and
time dilation are physical phenomena which affect bodies in motion
relative to the luminiferous ether. While they solved problems
relating to Electrodynamics, the idea of a length contraction due to
motion seemed incompatible with the notion of a rigid body being
rigid. The idea seemed to be introduced ad hoc and could not be
derived from any deeper insight lending it plausibility.  This is in
stark contrast to Albert Einstein’s elegant theories of relativity,
derived from the postulate of a constant speed of light. In special
relativity (SR), the Lorentz transformation is used to relate two
inertial frames \citep{Einstein1905}. Although mathematically
equivalent to SR, the LET was dismissed by the scientific
community. Nowadays, LET is not well known and many physicists believe
that the Michelson-Morley experiment conducted in 1887 at the Case
Western Reserve University in Cleveland, Ohio, disproved the existence
of the luminiferous ether \citep{MM}. This rejection is based on the
concept that the measurement devices are rigid bodies, unaffected by
their movement through the luminiferous ether.  However, if particles
are waves, then this may not be a correct description
\citep{schmidkroupa2014}.

Mathematically, the theories of LET and SR are equivalent, the differences seem to be on the philosophical side. The discussion took place before the advent of quantum mechanics. Today, it is generally accepted that matter has (at least partially, with the wave-particle dualism still not being properly understood) a wave nature. If the matter travelling through an ether is not assumed to be a rigid body, but rather a wave, Lorentz transformations show up naturally and need not be introduced ad hoc. The philosophical differences may then be resolved in a dualism of perspectives, with LET representing the outside perspective of classical physics, while SR describes the same phenomena from the perspective of an observer made of and restricted to matter, including all measurement tools.

In that sense, this contribution may be seen as a first step into exploring a general LET, which includes the dynamics of SCOs, in analogy to the role of general relativity in relation to special relativity. But the mathematics works completely without a philosophical background.  
In Sec.~\ref{sec:results} below, it will be shown that the dynamics of SCOs exhibit general relativistic properties on a completely classical basis in Newtonian space and time, which is of interest in its own right and may provide, if nothing more, an educational perspective.

	
\section{Methods}
\label{sec:methods}

This section presents additional important concepts and tools derived here. Sec.~\ref{sec:basicknowledge} presents the required knowledge about the superposition of plane waves and Lorentz transformations. As already mentioned, there is an interesting connection between these two concepts. The following subsections show that any actively Lorentz transformed oscillation solves the classical wave equation, see Sec.~\ref{subsec:activelorentz}. In Sec.~\ref{subsec:visualisation}, a visualisation method is presented to help the reader imagine the effect of a Lorentz transformation on an arbitrary oscillation. Next, in Sec.~\ref{subsec:wavevector}, it is shown that Lorentz transformations emerge naturally in an inhomogeneous medium. Using plane wave superpositions, it is described how this affects oscillations as a whole in Sec.~\ref{subsec:planewavesolution}.

\subsection{Active Lorentz Transformation of an Oscillation}
\label{subsec:activelorentz}

For an oscillation to remain a solution of the wave equation after an active Lorentz transformation, it is sufficient to show that any plane wave conserves the form of a plane wave under the same transformation. Any oscillation can be decomposed into plane waves. If each plane wave is actively transformed into a different plane wave, the transformed oscillation can still be written as a superposition of plane waves, thus solving the wave equation. For a transformation to conserve the form of a plane wave, the expression in the complex exponential function has to behave as follows:
\begin{equation}\label{conserveform}
\mathbf{k}\cdot\mathbf{x}' - kc_st' = \mathbf{k}'\cdot\mathbf{x} - k'c_st.
\end{equation}

After inserting $\mathbf{x}'$ and $t'$, it is required that the expression can be rewritten with $\mathbf{k}'$ and $k' = |\mathbf{k}'|$, which are the properties of the transformed plane wave. If this is not possible, the transformation does not conserve the form of a plane wave. To analyse if a Lorentz transformation conserves the form of a plane wave, the equations given in Eq.\ \ref{Lorentz3} and Eq.\ \ref{Lorentz32} are inserted in Eq.\ \ref{conserveform}:
\begin{align}
\mathbf{k}\cdot\mathbf{x}' - kc_st' ={}& \mathbf{k}\cdot[\gamma[\mathbf{x}\cdot\mathbf{n} - \beta ct]\mathbf{n} + \mathbf{x} - \mathbf{n}(\mathbf{x}\cdot\mathbf{n})] \nonumber\\
& - k[\gamma c_st -\beta\gamma\mathbf{n}\cdot\mathbf{x}] \nonumber\\
={}& [\gamma[\mathbf{k}\cdot\mathbf{n} + \beta k]\mathbf{n} + \mathbf{k} - \mathbf{n}(\mathbf{k}\cdot\mathbf{n})]\cdot\mathbf{x} \nonumber\\
& - [k\gamma + \beta\gamma\mathbf{n}\cdot\mathbf{k}]c_st \nonumber\\
={}& \mathbf{k}'\cdot\mathbf{x} - k'c_st.
\end{align}

This calculation is not sufficient to prove that a Lorentz transformation conserves the form of a plane wave, as the condition $k' = |\mathbf{k}'|$ has to be fulfilled at the same time as
\begin{align}
|\mathbf{k}'| ={}& \sqrt{|\mathbf{k}'\times\mathbf{n}|^2+(\mathbf{k}'\cdot\mathbf{n})^2} \nonumber\\
={}& \sqrt{|\mathbf{k}\times\mathbf{n}|^2 + \gamma^2(\mathbf{k}\cdot\mathbf{n}+\beta k)^2} \nonumber \\
={}& \sqrt{\gamma^2k^2 + 2\gamma^2 k\beta\mathbf{n}\cdot\mathbf{k} + \beta^2\gamma^2(\mathbf{n}\cdot\mathbf{k})^2} \nonumber\\
={}& \gamma[k+\beta\mathbf{n}\cdot\mathbf{k}] 
= k',
\end{align}
using the relation given in Eq.\ \ref{betagamma}.

These two calculations show that an actively Lorentz transformed plane wave results in another plane wave with properties $\mathbf{k}'$ and $k'$. The transformed wave vector and wave number are
\begin{align}
\mathbf{k}' ={}& \gamma[\mathbf{k}\cdot\mathbf{n} + \beta k]\mathbf{n} + \mathbf{k} - \mathbf{n}(\mathbf{k}\cdot\mathbf{n}) \\
& \Rightarrow 
\begin{cases}
\mathbf{k}'\times\mathbf{n} = \mathbf{k}\times\mathbf{n}, \\
\mathbf{k}'\cdot\mathbf{n} = \gamma[\mathbf{k}\cdot\mathbf{n} + \beta k],
\end{cases}\\
k' ={}& \gamma[k + \beta\mathbf{k}\cdot\mathbf{n}].\label{magkprime}
\end{align}

These equations for the new wave vector and wave number are similar to Eq.\ \ref{Lorentz3} and Eq.\ \ref{Lorentz32}. The wave number assumes the role of the time coordinate and the direction of the boost is reversed; i.e. $\mathbf{n}\to - \mathbf{n}$. It can be written as a matrix as
\begin{equation}
\begin{pmatrix}
k' \\ \mathbf{k}'
\end{pmatrix}
=
\begin{pmatrix}
\gamma & \beta\gamma\mathbf{n}^T \\
\beta\gamma\mathbf{n} & \mathbf{I} + (\gamma-1)\mathbf{n}\mathbf{n}^T
\end{pmatrix}
\begin{pmatrix}
k \\ \mathbf{k}
\end{pmatrix}.
\end{equation}

Thus, every solution of the wave equation can be actively Lorentz transformed and remains a solution of the wave equation. This is the connection between plane wave superpositions and Lorentz transformations. This procedure can be used to find related solutions of the wave equation; oscillations that are related by Lorentz transformations.

\subsection{Visualisation of Transformed Oscillations}
\label{subsec:visualisation}

The previous subsection shows that an active Lorentz transformation of an oscillation remains a solution of the wave equation. An active transformation in spacetime can be thought of as a mapping between events. An event is a point in space at a fixed time. An {\it active} Lorentz transformation relates the transformed and the not-transformed oscillation. It is a mathematical procedure acting on scalar functions and does not represent a physical change of spacetime. The effect of an active Lorentz transformation on an oscillation can schematically be described by the matrix
\begin{equation}\label{transmatrix}
\begin{pmatrix}
c_st' \\ \mathbf{x}'
\end{pmatrix} =
\begin{pmatrix}
\text{oscillation frequency}  & \text{phase shift} \\
\text{movement} & \text{distortion}
\end{pmatrix}
\begin{pmatrix}
c_st \\ \mathbf{x}
\end{pmatrix}.
\end{equation}

A factor relating the new and old time coordinate causes a change in the oscillation frequency at each point in space. A new time coordinate depending on old space coordinates represents a space-dependent phase shift. If a new space coordinate depends on time, each point of the oscillation moves through space. The matrix relating new and old space coordinates describes a distortion of the oscillation in space. Next, it is described how an oscillation as a whole changes if each of its plane wave constituents is actively Lorentz transformed.

\subsubsection{Transforming an Oscillation at Rest}
\label{subsec:oscatrest}

An oscillation with a well defined center at rest, without loss of generality at the origin, is denoted as $\rho_\mathrm{rest}(\mathbf{x},t)$. It is actively Lorentz transformed, which results in a related oscillation called $\rho_\mathrm{Lor}(\mathbf{x},t)$. Each oscillation is a scalar function in spacetime. They are related by an active Lorentz transformation which can be written as
\begin{multline}\label{restrel}
\rho_\mathrm{Lor}(\mathbf{x},t) = \\ \rho_\mathrm{rest}\left(\gamma[\mathbf{x}\cdot\mathbf{n} - \beta c_st]\mathbf{n} + \mathbf{x} - \mathbf{n}(\mathbf{x}\cdot\mathbf{n}),\gamma\left[t - \frac{\beta(\mathbf{x}\cdot\mathbf{n})}{c_s}\right]\right).
\end{multline}
 
This equation contains all information how an oscillation at rest is affected by an active Lorentz transformation. The interpretation of Eq.\ \ref{restrel} can be separated into a few steps. The center of the transformed oscillation, denoted as $\mathbf{x}_c(t)$, is followed through space, allowing a comparison of the oscillations in their relative resting frames. The center of the oscillation at rest $\rho_\mathrm{rest}$ is located at the origin. The center of the transformed oscillation is determined by setting the space-dependence of $\rho_\mathrm{rest}$ to zero:
\begin{align}
\gamma[\mathbf{x}_c(t)\cdot\mathbf{n} - \beta c_st]\mathbf{n} + \mathbf{x}_c(t) - \mathbf{n}(\mathbf{x}_c(t)\cdot\mathbf{n}) = 0 \nonumber\\
\Rightarrow \mathbf{x}_c(t) = \beta c_s t\mathbf{n}.
\end{align}

The center of the transformed SCO moves at velocity $\beta c_s$ in the $\mathbf{n}$-direction. The oscillation frequencies at the center of the oscillation at rest and the center of the transformed oscillation are different. This can be seen by inserting the position of the transformed center into Eq.\ \ref{restrel},
\begin{equation}\label{lorproptime}
\rho_\mathrm{Lor}(\beta c_s t\mathbf{n},t) = \rho_\mathrm{rest}\left(0,\frac{t}{\gamma}\right).
\end{equation}

The mapping of the values at the center of the oscillation at rest to the center of the transformed oscillation undergoes slowdown. A time period $T_\mathrm{rest}$ at rest is performed in a longer time period $T_\mathrm{motion} =\gamma T_\mathrm{rest}$ in motion.

Introducing the distance from the transformed center as $\mathbf{x}_{cd} = \mathbf{x} - \beta c_st\mathbf{n}\;$ allows the comparison of the oscillations in their respective resting frames as
\begin{multline}\label{restframe}
\rho_\mathrm{Lor}(\beta c_s t\mathbf{n} + \mathbf{x}_{cd},t) = \\\rho_\mathrm{rest}\left(\mathbf{x}_{cd} - (\gamma-1)\mathbf{n}(\mathbf{x}_{cd}\cdot\mathbf{n}),\frac{t}{\gamma} - \frac{\gamma\beta(\mathbf{x}_{cd}\cdot\mathbf{n})}{c_s}\right).
\end{multline}

\begin{figure}
\centering
\includegraphics[width=0.33\columnwidth, angle=-90]{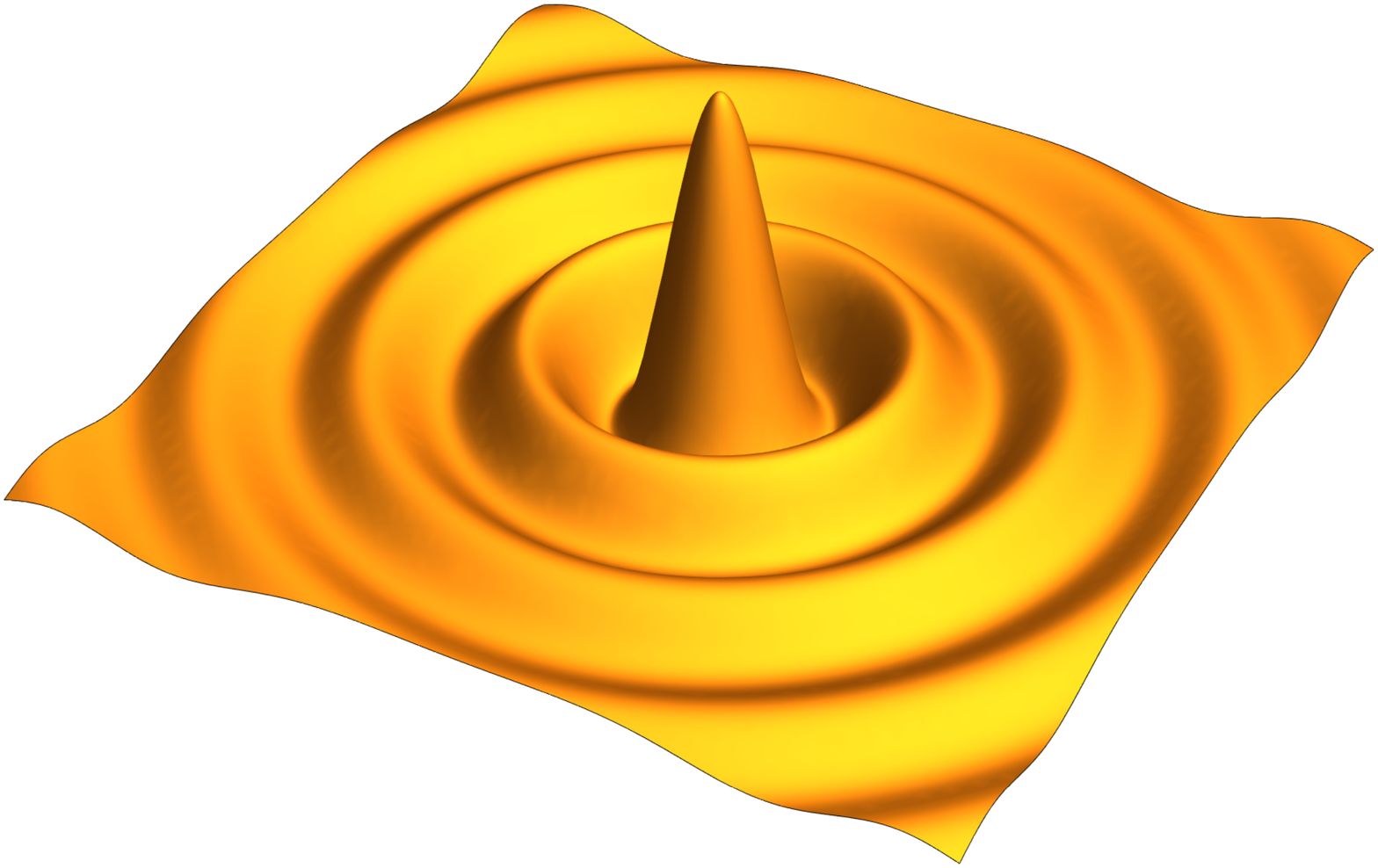}
\caption{Two-dimensional radially symmetric oscillation at rest at a fixed time.}
\label{png2Dosc1}
\end{figure}

\begin{figure} 
\centering
\includegraphics[width=0.33\columnwidth, angle=-90]{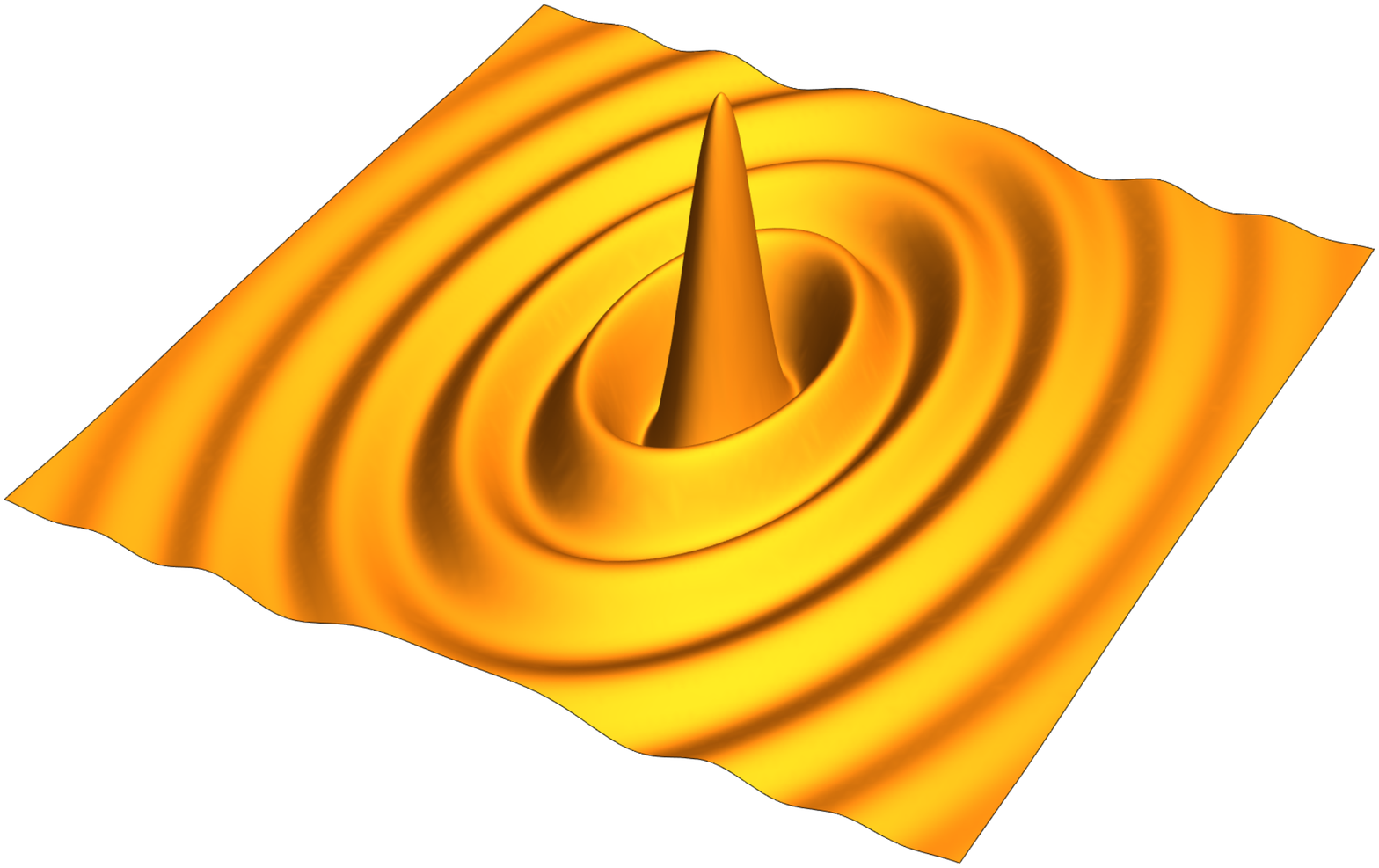} 
\caption{Two-dimensional radially symmetric oscillation after active Lorentz transformation at a fixed time as viewed from the reference frame of the not transformed oscillation.}
\label{png2Dosc2}
\end{figure}

The term $\beta c_s t\mathbf{n} + \mathbf{x}_{cd}$ in the space dependence of $\rho_\mathrm{Lor}$ describes the values of the transformed oscillation at the distance $\mathbf{x}_{cd}$ from the transformed center. These values originate from the oscillation at rest as described by the right hand side of Eq.\ \ref{restframe}. At time $t=0$, the shape of the transformed oscillation is described by the values of the not transformed oscillation found at the position $\mathbf{x}_{cd} - (\gamma-1)\mathbf{n}(\mathbf{x}_{cd}\cdot\mathbf{n})$ at time $t = - \gamma\beta(\mathbf{x}_{cd}\cdot\mathbf{n})/c_s$. This represents a contraction of the oscillation in the $\mathbf{n}$-direction, accompanied by a space-dependent phase shift. Additionally, the overall oscillation frequency is changed by the factor $1/\gamma$. This is the effect of an active Lorentz transformation on an oscillation at rest. See Fig.\ \ref{png2Dosc1} and Fig.\ \ref{png2Dosc2} for the visualisation of the Lorentz contraction on a two-dimensional radially symmetric oscillation.

\subsubsection{Transforming an Oscillation in Motion}
\label{subsec:oscinmotion}

An oscillation in motion can be constructed by an active Lorentz transformation of an oscillation at rest, as shown in the previous subsection. The initial oscillation in motion is thus given by $\rho_\mathrm{Lor}(\mathbf{x},t)$ as given in Eq.\ \ref{restrel}. This initial oscillation is Lorentz transformed a second time in a different direction with a different parameter. The combined effect of two non-collinear Lorentz transformation is described by Thomas-Wigner rotation. Using Eq.\ \ref{Wigtwocases2}, the active transformation between the oscillation at rest $\rho_\mathrm{rest}$ and the Thomas-Wigner transformed oscillation $\rho_\mathrm{Wig}$ can be written as
\begin{multline}\label{motionrel}
\rho_\mathrm{Wig}(\mathbf{x},t) = \\\rho_\mathrm{rest}\left(\mathbf{R}(\vec{\alpha})[\gamma(\mathbf{x}\cdot\mathbf{p}-\beta c_st)\mathbf{p} + \mathbf{x} -\mathbf{p}(\mathbf{x}\cdot\mathbf{p})],
 \gamma\left[t - \frac{\beta(\mathbf{x}\cdot\mathbf{p})}{c_s}\right]\right).
\end{multline}

Following the same procedure, the center of the transformed oscillation can be found at the position at which the space dependence of $\rho_\mathrm{rest}$ is zero:
\begin{align}
\mathbf{R}(\vec{\alpha})[\gamma(\mathbf{x}_c(t)\cdot\mathbf{p}-\beta c_st)\mathbf{p} + \mathbf{x}_c(t) -\mathbf{p}(\mathbf{x}_c(t)\cdot\mathbf{p})] = 0 \nonumber\\
\Rightarrow \mathbf{x}_c(t) = \beta c_st\mathbf{p}.
\end{align}

The rotation matrix has no effect on the movement of the center of the transformed oscillation. The transformed oscillation moves with velocity $\beta c_s$ in the $\mathbf{p}$-direction. The active time transformation between the center of the oscillation at rest and the center of the transformed oscillation is
\begin{equation}\label{wigproptime}
\rho_\mathrm{Wig}(\beta c_st\mathbf{p},t) = \rho_\mathrm{rest}\left(0,\frac{t}{\gamma}\right).
\end{equation}

This is the same result as found in the previous subsection, describing a slowdown. As the last step, the distance from the transformed center is introduced as $\mathbf{x}_{cd} = \mathbf{x} - \beta c_st\mathbf{p}$. This allows the comparison of the oscillations in their relative resting frames,
\begin{multline}
\rho_\mathrm{Wig}(\beta c_s t\mathbf{p} + \mathbf{x}_{cd},t) = \\ \rho_\mathrm{rest}\left(\mathbf{R}(\vec{\alpha})[\mathbf{x}_{cd} - (\gamma-1)\mathbf{p}(\mathbf{x}_{cd}\cdot\mathbf{p})],\frac{t}{\gamma} - \frac{\gamma\beta(\mathbf{x}_{cd}\cdot\mathbf{p})}{c_s}\right).
\end{multline}

The effect of the spatial dependence of $\rho_\mathrm{rest}$ can be described in two separate steps. The oscillation at rest is first rotated by $\mathbf{R}(-\vec{\alpha})$ and then spatially contracted in $\mathbf{p}$-direction. The space-dependent phase shift then accompanies the contraction in $\mathbf{p}$ direction. For a spherically symmetric oscillation the spatial rotation can be neglected and the shape of $\rho_\mathrm{Wig}$ is the same as the shape of $\rho_\mathrm{Lor}$, only orientated in a different direction. For all other oscillations, the shape of $\rho_\mathrm{Wig}$ is obtained by performing a spatial rotation by $\mathbf{R}(-\vec{\alpha})$ first, then performing a single Lorentz transformation in $\mathbf{p}$-direction.

\subsection{Wave Vector in a Speed of Propagation Field}
\label{subsec:wavevector}

The previous subsections show that an active Lorentz transformation of an oscillation results in another solution of the wave equation. Following that, the effect of an active Lorentz transformation on an arbitrary oscillation is explained. Until now, these active transformations are only a mathematical procedure, not a physical phenomenon. The following subsection shows that the Lorentz transformation naturally emerges in plane waves travelling through inhomogeneous mediums. 

Is there a physical process which affects plane waves? The most commonly known effect is the refraction of a plane wave between two mediums with different speeds of propagation, described by the Ibn-Sahl--Snell's law. On the boundary between the two mediums, the oscillation has to be continuous which demands the same value for the oscillation frequency in both mediums. Due to the different speeds of propagation, the wave number has to change accordingly. A change in direction of the plane wave can be derived via Fermat's principle of least travel time or Huygens' principle as presented in basic physics textbooks (e.g. \cite{gerthsen2010gerthsen}). Due to the infinitely extended nature of the plane waves, it is not possible to straightforwardly apply the same mechanism to continuous space-dependent speed of propagation fields.

The oscillations of interest have to be limited to SCOs. This makes it
possible to derive the behaviour of a single plane wave, as a
constituent of an SCO, in an arbitrary space-dependent speed of propagation
field. The resulting law is referred to here as the continuous Ibn-Sahl--Snell's law.

\subsubsection{Continuous Ibn-Sahl--Snell's Law of Refraction}
\begin{figure}
\includegraphics[width=\columnwidth]{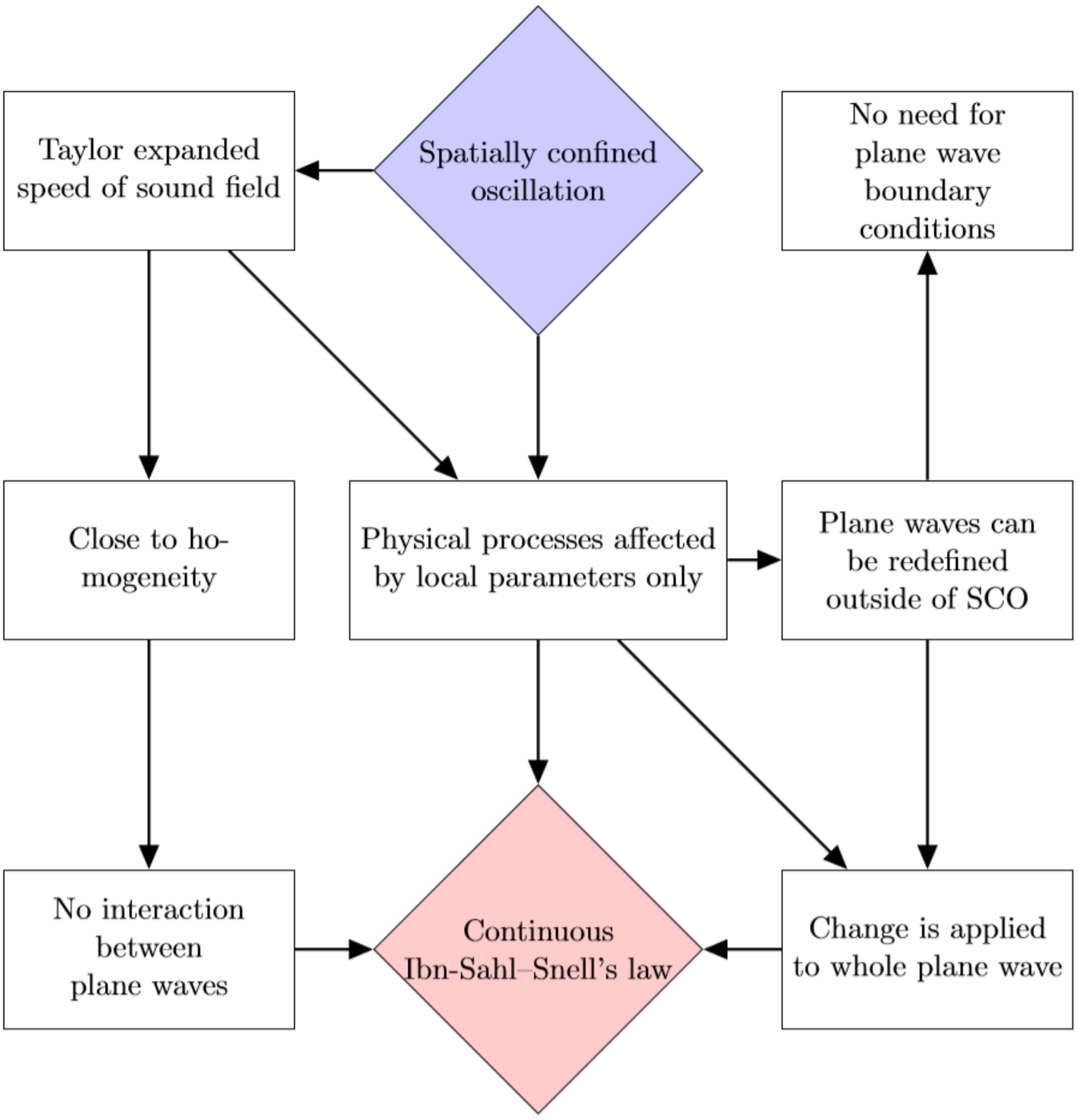}
\vspace{-2cm}
\caption{Line of reasoning to justify usage of continuous Ibn-Sahl--Snell's law for plane wave constituents of SCOs in inhomogeneous mediums.}
 \label{pngwaveinhomo}
\end{figure}

This section presents, in preparation of the derivation, a line of reasoning to arrive at the conclusion that the change of a plane wave, as a constituent of an SCO, is described by a continuous Ibn-Sahl--Snell's law. The line of reasoning is summarised in Fig.\ \ref{pngwaveinhomo}. This figure can be used alongside the explanation to facilitate the understanding. Each of the following paragraphs is represented by a field in Fig.\ \ref{pngwaveinhomo}.

\paragraph{Spatially confined oscillation.}
As the only assumption, an SCO is positioned, without loss of generality, at the origin of the coordinate system in an arbitrary inhomogeneous speed of propagation field $c_s(\mathbf{x})$, see upper central entry in Fig.\ \ref{pngwaveinhomo}. 

\paragraph{Taylor expand speed of propagation field.}
In the vicinity of the SCO, the speed of propagation field $c_s(\mathbf{x})$ can be Taylor expanded to first order, see Fig.\ \ref{pngwaveinhomo}, upper left entry:
\begin{align}\label{taylorfield}
c_s(\mathbf{x}) ={}& c_s(0) +  \mathbf{\nabla}c_s(\mathbf{x})|_{\mathbf{x}=\mathbf{0}}\cdot \mathbf{x} + \mathcal{O}(|\mathbf{x}|^2) \nonumber\\
={}& c_{s,0} + \mathbf{c}_s'\cdot\mathbf{x} + \mathcal{O}(|\mathbf{x}|^2).
\end{align}

The value of the speed of propagation field at the origin is denoted as $c_{s,0}$. The spatial gradient of the speed of propagation field at the origin is called $\mathbf{c}_s'$ with magnitude $c_s'$ and direction $\mathbf{e}'$. The SCO has to be much smaller than the length scale $L=c_s'/\mathrm{maxeigen}(\mathbf{H})$, with $\mathrm{maxeigen}(\mathbf{H})$ being the largest eigenvalue of the Hessian matrix of $c_s(\mathbf{x})$. The ratio $L$ describes how fast the magnitude of the spatial gradient changes in space. For an SCO much smaller than $L$, the spatial gradient can be assumed constant on the scale of the SCO.

\paragraph{Physical processes affected by local parameters only.}
The Taylor expanded speed of propagation field changes linearly in space, as described by its local parameters. Due to the spatial confinement of the SCO, all physical processes (e.g.\ the interaction of the oscillation with the inhomogeneous medium) are only affected by these local parameters, see central entry in Fig.\ \ref{pngwaveinhomo}. 

\paragraph{Plane waves can be redefined outside of the SCO.}
As the next step, the SCO is decomposed into infinitely extended plane waves. These plane waves are physically relevant only in the vicinity of the SCO. The parts of the plane waves outside of the SCO can effectively be redefined in any way, as long as their interference adds up to zero, see central right entry in Fig.\ \ref{pngwaveinhomo}. 

\paragraph{Change is applied to the whole plane wave.}
The possible redefinition of the infinitely extended plane waves outside of the SCO allows the physical changes of the plane waves in the vicinity of the SCO to be applied to the infinitely extended plane waves (see lower right entry of Fig.\ \ref{pngwaveinhomo}). This is only possible if the changing plane waves remain plane waves, e.g.\ no curving of wave fronts. This has to be fulfilled at all times. The redefinition also has to fulfil the conditions of the interference of all plane waves still adding up to zero outside of the SCO. This condition is fulfilled if the SCO remains spatially confined while being affected by the inhomogeneity. This has to be checked later.

\paragraph{No need for plane wave boundary conditions.}
The boundary condition that has to be fulfilled is the spatial confinement of the SCO; the SCO has to effectively reach zero at sufficiently large distances from its center. This condition has to be fulfilled at all times. There is no need for boundary conditions for the infinitely extended plane waves in this case, see upper right entry in Fig.\ \ref{pngwaveinhomo}.

\paragraph{Close to homogeneity.}
Going back to the Taylor expanded speed of propagation field, the spatial gradient is small, in the sense that the speed of propagation changes by a small amount on the scale of the SCO. For such small spatial gradients, equivalent to the homogeneous case, the plane waves do not interact with each other, see lower left entry of Fig.\ \ref{pngwaveinhomo}. Individually, they respond to the change in the speed of propagation field in the surroundings of the oscillation; the plane waves react to the spatial gradient in the speed of propagation field.  

\paragraph{Continuous Ibn-Sahl--Snell's law}
The line of reasoning above leads to the continuous version of the Ibn-Sahl--Snell's law, see lower central entry of Fig.\ \ref{pngwaveinhomo}. There are three reasons which motivate the usage of the continuous Ibn-Sahl--Snell's law. First, due to the closeness to homogeneity, each plane wave constituent of an SCO can be described individually. Second, the plane wave is only considered in the vicinity of the SCO in which the spatial gradient is constant. Third, the change is applied to the whole plane wave, with the condition of all plane waves adding up to zero outside of the SCO.

Similar to the discrete version, the continuous version of the Ibn-Sahl--Snell's law can be derived in different ways, all leading to the same result. One of these ways is to use wave front segments travelling at different speed of propagations, leading to a change in wave number and direction of motion of the plane wave. This approach is used in the following subsections to derive the continuous version of the Ibn-Sahl--Snell's law. 

\begin{figure}
\centering
\includegraphics[width=0.33\textwidth,angle=-90]{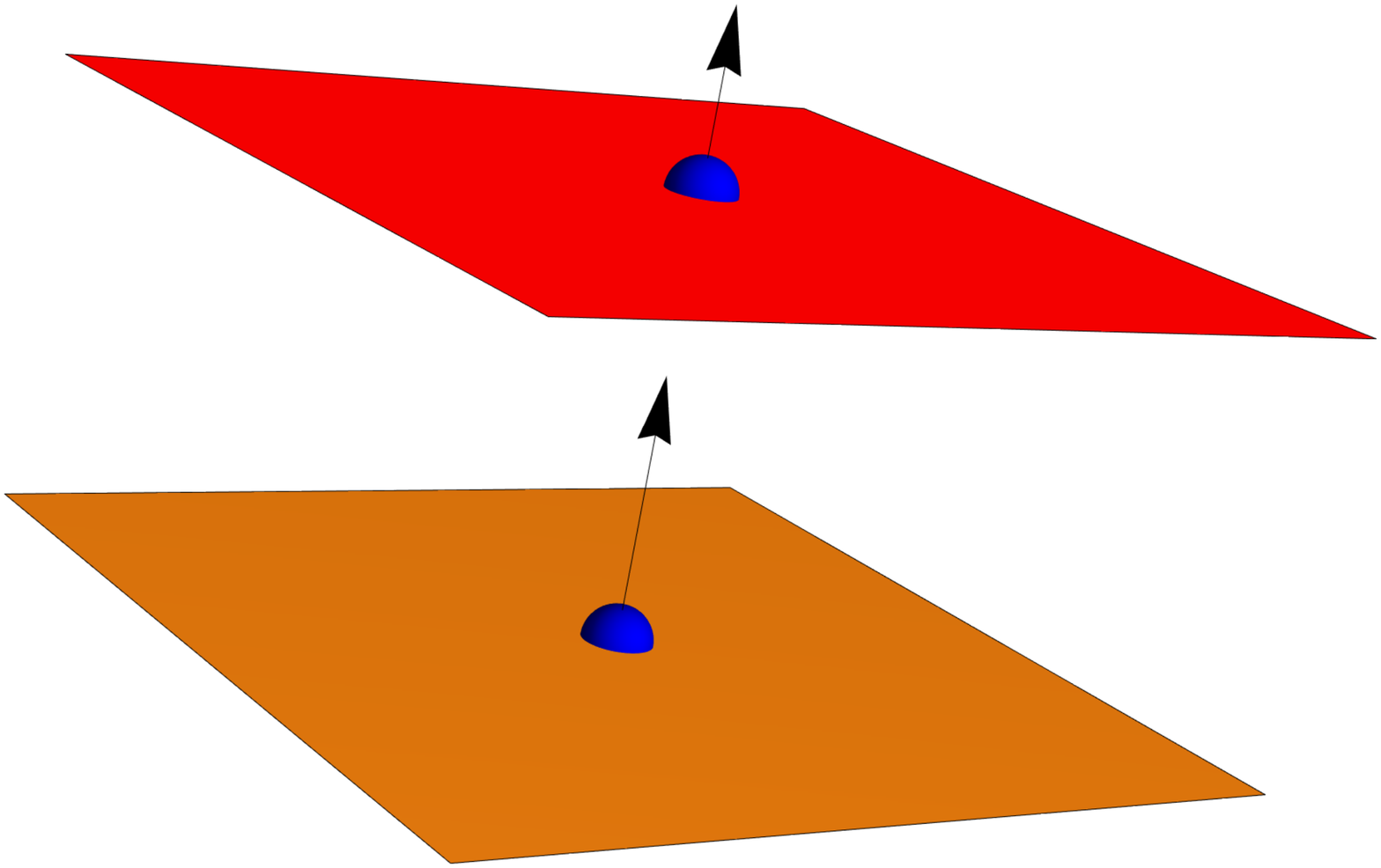}
\caption{Two corresponding wave segments of two different wave fronts traveling at different velocities.}
\label{pngcorreswavseg}
\end{figure}

\subsubsection{Change in Wave Number}

Two subsequent wave fronts have corresponding wave front segments, which are defined by the minimal distance between them (see Fig.\ \ref{pngcorreswavseg}). These two wave front segments may have different propagation velocities due to their different position in space. Following the two wave front segments for a time step $\mathrm{d}t$ reveals a change in their distance which translates to a change in the overall wave length and wave number between the wave fronts, given that neighbouring pairs of segments behave the same. The movement of the two corresponding wave front segments is summarised in Table \ref{tabwavelength}. The difference between the positions at time $t+\mathrm{d}t$,
\begin{equation}\label{wavelengthdt}
\Delta = \lambda + \lambda\mathbf{c}'_s\cdot\mathbf{s}\mathrm{d}t,
\end{equation}
is independent of the position of the wave front segments in space. All pairs of corresponding wave front segments in two subsequent wave fronts thus behave the same way and the length $\Delta$ can be identified as the wavelength after a time step $\mathrm{d}t$, $\lambda(t+\mathrm{d}t)$. This leads to the differential equation for the time-rate of change of the wavelength as
\begin{equation}
\dot{\lambda}(t) = \mathbf{c}'_s\cdot\mathbf{s}\lambda(t),
\end{equation}
or, with $k=2\pi/\lambda$ as the time-rate of change of the wave number as
\begin{equation}\label{kchange}
\dot{k}(t) = - \mathbf{k}(t)\cdot\mathbf{c}'_s.
\end{equation}

This differential equation only depends on the spatial gradient which is constant in the region of the SCO. The position of the pair of corresponding wave front segments is arbitrary. Each plane wave, as a whole, changes its wave number according to Eq.\ \ref{kchange}. If the wave fronts remain planes depends on the relative movement of multiple wave front segments in one wave front.

\begin{center}
\begin{table*}
\caption{Summary of the movement of two corresponding wave front
  segments of subsequent wave fronts.} 
\label{tabwavelength}
\begin{tabular}{ccc}
\hline \hline
property & segment 1 & segment 2 \\
\hline
position at time $t$ & $\mathbf{x}$ & $\mathbf{x} + \lambda\mathbf{s}$ \\
propagation velocity & $c_{s,0} + \mathbf{c}_s'\cdot\mathbf{x}$ & $c_{s,0} + \mathbf{c}_s'\cdot[\mathbf{x} + \lambda\mathbf{s}]$ \\
position at time $t+\mathrm{d}t$ & $\mathbf{x} + [c_{s,0} + \mathbf{c}_s'\cdot\mathbf{x}]\mathbf{s}\mathrm{d}t$ & $\mathbf{x}_1 + [c_{s,0} + \mathbf{c}_s'\cdot[\mathbf{x} + \lambda\mathbf{s}]\mathbf{s}\mathrm{d}t$ \\
\hline \hline
\end{tabular}
\end{table*}
\end{center}

\begin{center}
\begin{table*}
\caption{Summary of the relative movement of two neighbouring wave front segments with respect to segment positioned at $\mathbf{x}$.} 
\label{tabdirection}
\begin{tabular}{ccc}
\hline \hline
property & segment 2 & segment 3 \\
\hline
relative position at time $t$ & $\frac{\mathbf{s}\times\mathbf{e}'}{|\mathbf{s}\times\mathbf{e}'|}\mathrm{d}l$ & $\frac{\mathbf{s}\times(\mathbf{e}'\times\mathbf{s})}{|\mathbf{s}\times(\mathbf{e}'\times\mathbf{s})|}\mathrm{d}l$  \\
relative propagation velocity & 0 & $\mathbf{c}'_s\cdot\frac{\mathbf{s}\times(\mathbf{e}'\times\mathbf{s})}{|\mathbf{s}\times(\mathbf{e}'\times\mathbf{s})|}\mathrm{d}l$ \\
relative position at time $t+\mathrm{d}t$ & $\frac{\mathbf{s}\times\mathbf{e}'}{|\mathbf{s}\times\mathbf{e}'|}\mathrm{d}l$  & $\frac{\mathbf{s}\times(\mathbf{e}'\times\mathbf{s})}{|\mathbf{s}\times(\mathbf{e}'\times\mathbf{s})|}\mathrm{d}l + \mathbf{c}'_s\cdot\frac{\mathbf{s}\times(\mathbf{e}'\times\mathbf{s})}{|\mathbf{s}\times(\mathbf{e}'\times\mathbf{s})|}\mathrm{d}l\mathbf{s}\mathrm{d}t$ \\
\hline \hline
\end{tabular}
\end{table*}
\end{center}

\begin{figure}
\centering
\includegraphics[width=0.7\columnwidth, angle=-90]{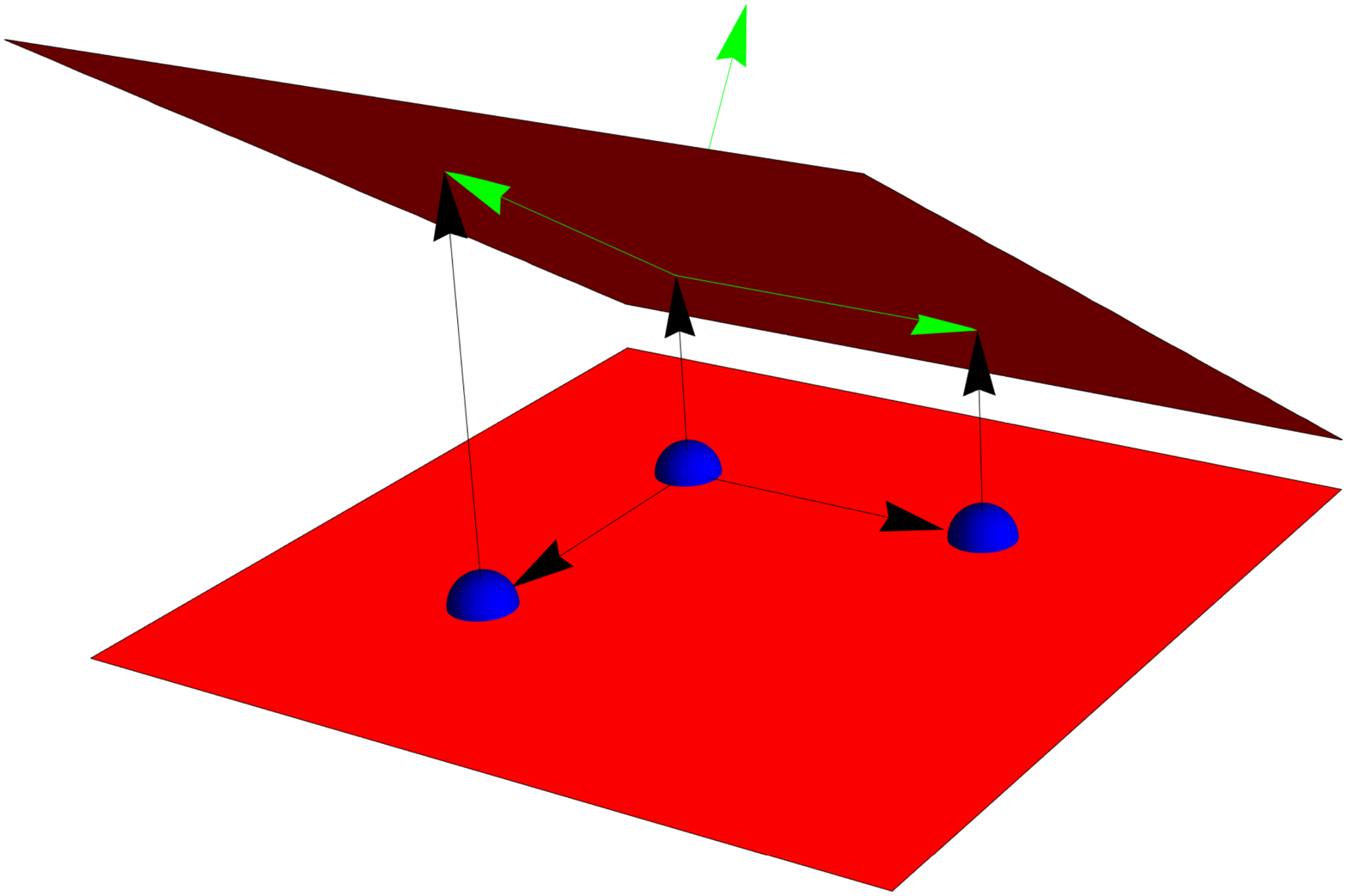}
\caption{Three wave front segments with one traveling at a different velocity, tilting the wave front plane.}
\label{pngthreewaveseg}
\end{figure}

\subsubsection{Change in Direction}

To derive the change in direction, three neighbouring wave front segments of a single wave front are followed for a time step $\mathrm{d}t$ (Fig.~\ref{pngthreewaveseg}). At time $t=0$, the three segments define a plane, with the normal vector $\mathbf{s}$, the direction of propagation of the wave front as a whole. In the case of $\mathbf{s}$ pointing in the direction of the spatial gradient, all segments have the same velocity of propagation and the direction of the wave front as a whole does not change, although the wavelength changes. The projection of $\mathbf{c}_s'$ on the plane is zero.

In any other case, the direction of $\mathbf{c}'_s$, referred to as $\mathbf{e}'$, can be used to define the initial positions of the three wave front segments.  The first segment is positioned at an arbitrary $\mathbf{x}$ with propagation velocity $c_{s,0} + \mathbf{c}'_s\cdot\mathbf{x}$ in direction $\mathbf{s}$. The second segment is positioned on the plane defined with normal vector $\mathbf{s}$, in the direction perpendicular to the projection of the spatial gradient at a distance $\mathrm{d}l$. It has the same propagation velocity as segment 1. The third segment is also positioned on the plane, the same length $\mathrm{d}l$ away from segment 1, but in the direction of the spatial gradient projection. It has a different propagation velocity from the other two segments, which leads to a tilting of the wave front plane as a whole. The relative movements of the segments 2 and 3 with respect to segment 1 are summarised in Table \ref{tabdirection}.

The relative position vectors given in Table \ref{tabdirection} connect the three wave front segments after a time step $\mathrm{d}t$. They can be used to find the new orientation of the wave front plane. The normal vector on the wave front plane after a time step $\mathrm{d}t$ is given by the normalised cross product of the two relative position vectors. The length $\mathrm{d}l$ cancels. The larger $\mathrm{d}l$, the more different the relative propagation velocity of segment 3, but the same amount of tilting is achieved because segment 3 has to traverse a larger distance to cause the same amount of tilting. The normalised cross product of the relative position vectors of the segments 2 and 3 is: 
\begin{align}\nonumber
\mathbf{s}(t+\mathrm{d}t) ={}& \text{Norm}\left[\left[\frac{\mathbf{s}\times\mathbf{e}'}{|\mathbf{s}\times\mathbf{e}'|}\right] \!\times\! \left[\frac{\mathbf{s}\times(\mathbf{e}'\times\mathbf{s})}{|\mathbf{s}\times(\mathbf{e}'\times\mathbf{s})|} \right.\right. \nonumber\\
& \left.\left.+ \mathbf{c}'_s\cdot\frac{\mathbf{s}\times(\mathbf{e}'\times\mathbf{s})}{|\mathbf{s}\times(\mathbf{e}'\times\mathbf{s})|}\mathbf{s}\mathrm{d}t \right]\right] \nonumber\\ 
 ={}& \mathbf{s}(t) -\mathbf{s}(t)\times[\mathbf{c}'_s\times\mathbf{s}(t)]\mathrm{d}t.
\end{align}

This results in the differential equation for the time-rate of change of the direction of a wave front plane in a speed of propagation field with constant spatial gradient,
\begin{equation}\label{schange}
\dot{\mathbf{s}}(t) = - \mathbf{s}(t)\times[\mathbf{c}_s' \times \mathbf{s}(t)].
\end{equation}

The differential equation is independent of the position of segment 1. All wave fronts on the scale of the SCO are described by Eq.\ \ref{schange}. Each wave front plane changes by the same amount, so each plane wave conserves its form and simply changes its direction as a whole.

\subsubsection{Change in Wave Vector --  the Continuous Ibn-Sahl--Snell's Law}

Combining Eq.\ \ref{kchange} and Eq.\ \ref{schange} results in the differential equation for the time-rate of change of the wave vector of a plane wave as a constituent of an SCO in an inhomogeneous speed of propagation field, named here the continuous Ibn-Sahl--Snell's law, given as 
\begin{equation}\label{veckchange}
\dot{\mathbf{k}}(t) = -k(t)\mathbf{c}_s'.
\end{equation}

This differential equation describes the change of the plane wave constituents of an SCO in an inhomogeneous speed of propagation field with spatial gradient $\mathbf{c}_s'$. This differential equation is of crucial importance, as it introduces the Lorentz transformation as a physical phenomenon. 

As expected, the differential equation does not depend on the value of the local speed of propagation. The wave vector changes in negative $\mathbf{c}_s'$-direction. There is no dependence on any angle between direction of travel and spatial gradient. The change in wave vector is faster for larger wave numbers and larger spatial gradient. For a medium with constant spatial gradient, the solutions for the wave vector and wave number, with initial wave vector $\mathbf{k}$, are
\begin{align}\label{vecksol}
\mathbf{k}_\mathrm{sol}(t) ={}& \gamma_\mathrm{sol}[\mathbf{k}\cdot\mathbf{e}'  - \beta_\mathrm{sol}k]\mathbf{e}' + \mathbf{k} - \mathbf{e}'(\mathbf{k}\cdot\mathbf{e}') \\
& \Rightarrow
\begin{cases}
\mathbf{k}_\mathrm{sol}(t)\times\mathbf{e}' = \mathbf{k}\times\mathbf{e'}, \\
\mathbf{k}_\mathrm{sol}(t)\cdot\mathbf{e}' = \gamma_\mathrm{sol}[\mathbf{k}\cdot\mathbf{e}'  - \beta_\mathrm{sol}k],
\end{cases}\\
k_\mathrm{sol}(t)  ={}& \gamma_\mathrm{sol}[k - \beta_\mathrm{sol}\mathbf{k}\cdot\mathbf{e}'],
\end{align}
with
\begin{align}
\gamma_\mathrm{sol} \equiv{}& \gamma_\mathrm{sol}(t) = \cosh(c_s't), \\
\beta_\mathrm{sol} \equiv{}& \beta_\mathrm{sol}(t) = \tanh(c_s't), \\
\gamma_\mathrm{sol}\beta_\mathrm{sol} \equiv{}& \gamma_\mathrm{sol}(t)\beta_\mathrm{sol}(t) = \sinh(c_s't).
\end{align}

Eq.\ \ref{vecksol} is equivalent to an active Lorentz transformation of a plane wave in $-\mathbf{e}'$-direction with parameter $\gamma_\mathrm{sol}$, as shown in Sec.\ \ref{subsec:activelorentz}. That $|\mathbf{k}_\mathrm{sol}(t)| = k_\mathrm{sol}(t)$ is shown in Sec.\ \ref{subsec:activelorentz} as well. The differential equations Eq.\ \ref{kchange} and Eq.\ \ref{veckchange} can be checked:
\begin{multline}
\dot{k}_\mathrm{sol}(t) = -\mathbf{k}_\mathrm{sol}(t)\cdot\mathbf{c}_s' \\\Rightarrow c_s'\gamma_\mathrm{sol}\beta_\mathrm{sol}k - c_s'\gamma_\mathrm{sol}\mathbf{k}\cdot\mathbf{e}' = -\gamma_\mathrm{sol}[\mathbf{k}\cdot\mathbf{e}'  - \beta_\mathrm{sol}k]c_s',
\end{multline}
\begin{multline}
\dot{\mathbf{k}}_\mathrm{sol}(t) = - k_\mathrm{sol}(t)\mathbf{c}_s'
\\\Rightarrow c_s'\gamma_\mathrm{sol}\beta_\mathrm{sol}\mathbf{k}\cdot\mathbf{e}' - c_s'\gamma_\mathrm{sol}k = -\gamma_\mathrm{sol}[k - \beta_\mathrm{sol}\mathbf{k}\cdot\mathbf{e}']\cdot\mathbf{c}_s'.
\end{multline}

The following relations have been used, which are derived by using the derivatives of the functions $\cosh$ and $\sinh$,
\begin{align}\label{solrelations}
\frac{\mathrm{d}}{\mathrm{d}t}\gamma_\mathrm{sol}(t) ={}& c_s'\gamma_\mathrm{sol}(t)\beta_\mathrm{sol}(t), \\
\frac{\mathrm{d}}{\mathrm{d}t}[\gamma_\mathrm{sol}(t)\beta_\mathrm{sol}(t)] ={}& c_s'\gamma_\mathrm{sol}(t).
\end{align}

The solution given in Eq.\ \ref{vecksol} can only be used for a speed of propagation field with constant spatial gradient. For a changing spatial gradient, it is only possible to use it for infinitesimal small time steps, as the SCO might change position in space. Nevertheless, it is still possible to derive the effect of an arbitrary space-dependent speed of propagation field on an SCO as a whole. This is done later in Sec.\ \ref{subsec:inhomogeneousmedium} by calculating the instantaneous change in an SCO. First the effect of a speed of propagation field with constant spatial gradient on an SCO is derived in Sec.\ \ref{subsec:planewavesolution}.

\subsubsection{Dimensionless Continuous Ibn-Sahl--Snell's Law}
\label{subsec:dimensionless}

It can be seen that the magnitude of the spatial gradient $c_s'$ only affects how fast the change in wave vectors proceeds. This is also found in the differential equations describing the time-rate of change of plane waves. In each of them, the time coordinate can be substituted by the dimensionless  parameter $a = c_s't$. A derivative w.r.t.\ $a$ is denoted by a subscript $a$. The differential equation for the wave vector in Eq.\ \ref{veckchange} can be written as
\begin{equation}\label{dimensionlessk}
\mathbf{k}_a(a) = -k(a)\mathbf{e}'.
\end{equation}

Plane wave constituents and thus all oscillations in different spatial gradients experience the same effect, while this effect manifests itself faster or slower depending on the magnitude of the spatial gradient.

\subsection{Superposition of Plane Wave Solutions}
\label{subsec:planewavesolution}

The previous subsections derive the solution of a plane wave, as a constituent of an SCO, in an inhomogeneous speed of propagation field with constant spatial gradient. The solution is an active continuous Lorentz transformation with $\gamma_\mathrm{sol}(t)=\cosh(c_s't)$ in direction $-\mathbf{e}'$. These plane wave solutions have to be superimposed to derive the effect on an SCO as a whole. 

As a prerequisite for the following derivation, it is assumed that an SCO in an inhomogeneous medium changes its shape but remains spatially confined with a well defined center. This assumption is confirmed to be correct later. This allows the definition of a path of the SCO, described by $\mathbf{\lambda}(t)$. The path is directly related to the plane waves, as it represents the region of their constructive interference. As the position of the SCO changes, the local values of the speed of propagation field change. The change of each plane wave is accounted for by inserting the wave vector and wave number solutions given in Eq.\ \ref{vecksol} into the complex exponential function of any superposition representing an SCO. At the same time, to fulfil the dispersion relation, the local value of the speed of propagation has to be used. Thus, the total effect is described by
\begin{multline}\label{planewaveeffect}
\exp[\mathrm{i}\mathbf{k}\cdot\mathbf{x}-\mathrm{i}c_{s,0}t] \\
\to \exp[\mathrm{i}\mathbf{k}_\mathrm{sol}(t)\cdot\mathbf{x}-\mathrm{i}k_\mathrm{sol}(t)c_s(\mathbf{\lambda}(t))t].
\end{multline}

Using the local value of the speed of propagation is correct; the part of each plane wave positioned at the center of the SCO at each moment in time propagates at the speed of propagation at that position. This results in a passive change in the frequency of each plane wave. This change is caused by the movement of the SCO in a changing speed of propagation field. Eq.\ \ref{planewaveeffect} represents the effect of a speed of propagation field with constant spatial gradient on any plane wave which is a constituent of a moving SCO. This term can be inserted into any SCO-superposition of plane waves to describe its temporal evolution. 

The superposition of an arbitrary SCO at rest or in motion at the origin can be written as Eq.\ \ref{eq:superposition}:
\begin{equation}\label{SCOatrest}
\rho_\mathrm{init}(\mathbf{x},t) = \operatorname{Re}\left[\int\!\mathrm{d}^3k \,\tilde{\rho}(\mathbf{k})\exp[\mathrm{i}\mathbf{k}\cdot\mathbf{x}-\mathrm{i}kc_{s,0} t]\right].
\end{equation}

The exact shape of the oscillation is defined by the complex frequency spectrum. To obtain the temporal evolution of the SCO, Eq.\ \ref{planewaveeffect} is inserted into Eq.\ \ref{SCOatrest}:
\begin{multline}
\rho_\mathrm{sol}(\mathbf{x},t) = \\ \operatorname{Re}\left[\int\!\mathrm{d}^3k \,\tilde{\rho}(\mathbf{k})\exp[\mathrm{i}\mathbf{k}_\mathrm{sol}(t)\cdot\mathbf{x}-\mathrm{i}k_\mathrm{sol}(t)c_s(\mathbf{\lambda}(t)) t]\right],
\end{multline}
with the same complex frequency spectrum $\tilde{\rho}(\mathbf{k})$, because amplitude and phase offset of each plane wave stay the same. What does $\rho_\mathrm{sol}(\mathbf{x},t)$ look like? The properties of the SCO solution can be found by establishing an active transformation between the initial SCO and the SCO solution, as in Sec.\ \ref{subsec:visualisation}:
\begin{equation}\label{SCOrel1}
\rho_\mathrm{sol}(\mathbf{x},t) = \rho_\mathrm{init}(\mathbf{x}'(\mathbf{x},t),t'(\mathbf{x},t)).
\end{equation}

To find the active transformation, the term in the exponential function in $\rho_\mathrm{sol}(\mathbf{x},t)$ can be rearranged, with $\gamma_\mathrm{sol} = \gamma_\mathrm{sol}(t) = \cosh(c_s't)$ and $\beta_\mathrm{sol} = \beta_\mathrm{sol}(t) =\tanh(c_s't)$ as before,
\begin{multline}
 [\gamma_\mathrm{sol}[\mathbf{k}\cdot\mathbf{e}'  - \beta_\mathrm{sol}k]\mathbf{e} + \mathbf{k} - 
 \mathbf{e}'(\mathbf{k}\cdot\mathbf{e}')]\cdot\mathbf{x}\\-[\gamma_\mathrm{sol}[k - \beta_\mathrm{sol}\mathbf{k}\cdot\mathbf{e}']]c_s(\mathbf{\lambda}(t))t
 \\= \mathbf{k}\cdot[\gamma_\mathrm{sol}[\mathbf{x}\cdot\mathbf{e}' + \beta_\mathrm{sol}c_s(\mathbf{\lambda}(t))t]\mathbf{e}' + \mathbf{x} - \mathbf{e}'(\mathbf{x}\cdot\mathbf{e}')] \\
 - kc_s(\mathbf{\lambda}(t))\gamma_\mathrm{sol}\left[t + \frac{\beta_\mathrm{sol}\mathbf{x}\cdot\mathbf{e}'}{c_s(\mathbf{\lambda}(t))}\right] = \mathbf{k}\cdot\mathbf{x}'(\mathbf{x},t) - kc_{s,0}\,t'(\mathbf{x},t).\nonumber
\end{multline}

%
%

The equations of this active transformation are
\begin{align}\label{cont1}
\mathbf{x}'(\mathbf{x},t) ={}& \gamma_\mathrm{sol}[\mathbf{x}\cdot\mathbf{e}' + \beta_\mathrm{sol}c_s(\mathbf{\lambda}(t))t]\mathbf{e}' + \mathbf{x} - \mathbf{e}'(\mathbf{x}\cdot\mathbf{e}')\\
& \Rightarrow
\begin{cases}
\mathbf{x}'(\mathbf{x},t) \times \mathbf{e}' = \mathbf{x}\times \mathbf{e}', \\
\mathbf{x}'(\mathbf{x},t) \cdot \mathbf{e}' = \gamma_\mathrm{sol}[\mathbf{x}\cdot\mathbf{e}' + \beta_\mathrm{sol}c_s(\mathbf{\lambda}(t))t],
\end{cases}\\
   t'(\mathbf{x},t) ={}& \gamma_\mathrm{sol}\frac{c_s(\mathbf{\lambda}(t))}{c_{s,0}}t + \frac{\gamma_\mathrm{sol}\beta_\mathrm{sol}\mathbf{e}'\cdot\mathbf{x}}{c_{s,0}}.\label{cont12}
\end{align}

These equations can also be written using the definition for $\mathbf{B}$ as introduced in Eq.\ \ref{Lor}:
\begin{equation}
\begin{pmatrix}
c_{s,0}t' \\ \mathbf{x}'
\end{pmatrix}
= 
\mathbf{B}(\gamma_\mathrm{sol},-\mathbf{e}')
\begin{pmatrix}
c_s(\mathbf{\lambda}(t))t \\ \mathbf{x}
\end{pmatrix}.
\end{equation}

The spatial part is equivalent to a Lorentz transformation in $-\mathbf{e}'$ direction. The time part of the active transformation reveals not only the terms from a Lorentz transformation in $-\mathbf{e}'$-direction, but also an additional factor $c_s(\mathbf{\lambda}(t))/c_{s,0}$. This factor affects the overall oscillation frequency (upper left entry of Eq.\ \ref{transmatrix}). This active transformation will be called a Lorentz-type transformation. A constant spatial gradient has the effect of an active continuous Lorentz-type transformation on any SCO, regardless of its state of motion.

The continuous Lorentz-type transformation contains information about the acceleration of an SCO. This is shown in Sec.\ \ref{subsec:inhomogeneousmedium}. First, the properties of SCOs in homogeneous mediums are discussed in Sec.\ \ref{subsec:constspeed}.

	
\section{Results and Discussion}
\label{sec:results}

The methods presented in Sec.\ \ref{sec:basicknowledge} and Sec.\ \ref{sec:methods} can be used to establish a theory for the dynamics of SCOs in homogeneous and inhomogeneous mediums. An SCO in a homogeneous medium is an extension to the spheron in the spheronic toy universe introduced by \cite{schmidkroupa2014}. Both, an SCO in a homogeneous medium and a spheron, show the Lorentz contraction and time dilation of particles. 

To derive the dynamics of SCOs, different concepts are introduced for SCOs in homogeneous mediums, based on \cite{schmidkroupa2014}. The concepts are extended to inhomogeneous mediums. A naturally emerging particle-field interaction is derived in Sec.~\ref{subsec:inhomogeneousmedium} in the form of an equation of motion. This equation of motion is equivalent to the geodesic equation for flat-space metrics, see Sec.\ \ref{subsec:geodesictrajectory}.

\subsection{Medium with Constant Speed of Propagation}
\label{subsec:constspeed}

As a first step, a homogeneous medium with speed of propagation $c_{s,0}$ is considered. In this section, some concepts are introduced which are based on the work of \cite{schmidkroupa2014}. They will be extended to be used in inhomogeneous mediums as well. The concepts include the definition of a particle-prototype, an SCO-observer and the concept of proper time.

\subsubsection{Particle-Prototype}

The particle-prototype is introduced in \cite{schmidkroupa2014} in the context of a propagating standing spherical wave, called spheron. Here it is extended to an arbitrary SCO. A spheron is only one example of an SCO. As discussed in \citep{schmidkroupa2014}, spherons, and more general SCOs, exhibit particle-like features. They are localised and contain energy. An SCO can be stationary or in motion. A propagating SCO is obtained by an active Lorentz transformation of a stationary SCO. SCOs related by a Lorentz(-type) transformation can be defined to represent the same particle in different states of motion. Each particle is fully described by the complex frequency spectrum it adapts at rest. A propagating SCO in a homogeneous medium remains propagating, a stationary SCO remains stationary. This is Newton's first axiom. A change in the state of motion of an SCO occurs if the effect of the medium is not perfectly described by the wave equation. One possibility is an inhomogeneous medium. The effects of an inhomogeneous medium on an SCO are completely described in Sec.\ \ref{subsec:inhomogeneousmedium}. 

\subsubsection{SCO-Observer}

An observer made of SCOs will be denoted as an SCO-observer. This concept is found in \cite{schmidkroupa2014} in the context of a hypothetical observer made of spherons. As explained in \cite{schmidkroupa2014}, an SCO-observer is unable to determine its relative motion through the medium, leading to the same mathematical theory as special relativity. The SCO-observer can use its own oscillations to define a clock. This leads to the concept of proper time.

\subsubsection{Proper Time}
\label{subsec:propertimeinconst}

The following explanation is based on \cite{schmidkroupa2014}. The concept of proper time of an SCO-observer can be introduced by defining a 'tick' of its clock. One 'tick' can be defined as the time span it takes for the center of an SCO to go from one extremum to its next, e.g.\ minimum to maximum or vice-versa.

In comparison, trying to find the proper time elapsing for a plane wave fails. Plane waves in a classical medium always travel at the speed of propagation. Using any part of a plane wave in the attempt to define a 'tick' fails. The chosen part does not oscillate and no proper time elapses. This is equivalent to electromagnetic waves in SR for which the proper time also does not change. For SCOs, the concept of proper time arises by introducing superpositions of plane waves. Superposing plane waves creates an oscillation which travels at velocities smaller than the speed of propagation. The faster the SCO travels, the longer one 'tick' and the less proper time elapses. It is possible to calculate how proper time depends on the velocity of the SCO.

One 'tick' of an SCO at rest can be used to define the coordinate time $t$. Stationary SCOs can easily be synchronised in homogeneous mediums. In Sec.\ \ref{subsec:oscatrest} (Eq.\ \ref{lorproptime}) and Sec.\ \ref{subsec:oscinmotion} (Eq.\ \ref{wigproptime}), it is shown that active time transformation between the center of an oscillation at rest (non-primed) and the center of an oscillation in motion (primed) is
\begin{equation}
t' = \frac{t}{\gamma}.
\end{equation}

For a 'tick' of the oscillation at rest with length $\mathrm{d}t$, the 'tick' of the moving oscillation can be described as
\begin{equation}
\mathrm{d}t' = \mathrm{d}\tau = \frac{\mathrm{d}t}{\gamma}.
\end{equation}

\subsubsection{Remarks on Homogeneous Mediums}

Using waves as particles is already common in quantum mechanics, and this simple assumption naturally leads to a theory equivalent to special relativity, explaining the result of the famous Michelson-Morley experiment in 1887. In \cite{schmidkroupa2014} it is shown that spherons have features found similarly in quantum mechanics. It can be assumed that these features can be extended to SCOs as well. To examine the relation to quantum mechanics in more detail, the effect of an SCO on the surrounding medium has to be described; the SCO cannot represent a test-particle in this case. Particle-particle interactions have to be considered. On small scales, considered in quantum mechanics, the properties of the medium have to be defined in more detail. 

The next section describes how an SCO is affected by an inhomogeneous speed of propagation field.

\subsection{Inhomogeneous Medium}
\label{subsec:inhomogeneousmedium}

As shown in \cite{schmidkroupa2014}, SCOs in a homogeneous medium lead to a theory equivalent to special relativity. Interestingly, an inhomogeneous medium leads to SCOs travelling on geodesic trajectories, as will be shown in this subsection. SCOs travelling in a homogeneous medium travel on straight trajectories, which are also geodesics.

It is shown in Sec.\ \ref{subsec:planewavesolution} that a constant spatial gradient in a speed of propagation field has the effect of an active continuous Lorentz-type transformation. An SCO at rest is simply transformed by this Lorentz-type transformation and starts moving through space. For an initial SCO in motion, Thomas-Wigner rotation has to be used. This is shown in the next subsection. 

The equations derived in Sec.\ \ref{subsec:oscinmotion} have to be changed slightly, because the active transformation relating the initial SCO in motion and the final SCO is not a pure Lorentz transformation, but a Lorentz-type transformation with an additional factor $c_s(\mathbf{\lambda}(t))/c_{s,0}$, see Eq.\ \ref{cont12}. All other results shown in Sec.\ \ref{subsec:oscinmotion} can be used.

All effects derived in this section for SCOs can be applied to an SCO-observer. For a more detailed understanding of the properties of an SCO-observer in inhomogeneous mediums, some gedankenexperiments could be performed. This is beyond the scope of this contribution.

\subsubsection{Equation of Motion for SCOs}

To derive the motion of an SCO in an inhomogeneous medium with constant spatial gradient, an initial moving SCO is used as introduced in Sec.\ \ref{subsec:oscatrest}. The initial SCO is related to an SCO at rest via 
\begin{equation}
\rho_\text{init}(\mathbf{x}',t') = \rho_\text{rest}(\mathbf{x}''(\mathbf{x}',t'),t''(\mathbf{x}',t')),
\end{equation}
with Eq.\ \ref{Lor}
\begin{equation}
\begin{pmatrix}
c_{s,0}t'' \\ \mathbf{x}''
\end{pmatrix}
= 
\mathbf{B}(\gamma_i,\mathbf{n})
\begin{pmatrix}
c_{s,0}t' \\ \mathbf{x}'
\end{pmatrix}.
\end{equation}

This results in an initial SCO in motion in the $\mathbf{n}$-direction with velocity $\beta_ic_{s,0}$. As found in Sec.\ \ref{subsec:planewavesolution}, the effect of an inhomogeneous speed of propagation field is described by a continuous Lorentz-type transformation described by $\gamma_\mathrm{sol}$ and $-\mathbf{e}'$. The resulting final SCO solution, called $\rho_\mathrm{sol}$ is related to the initial moving SCO by
\begin{equation}
\rho_\text{sol}(\mathbf{x},t) = \rho_\text{init}(\mathbf{x}'(\mathbf{x},t),t'(\mathbf{x},t)),
\end{equation}
with
\begin{equation}
\begin{pmatrix}
c_{s,0}t' \\ \mathbf{x}'
\end{pmatrix}
= 
\mathbf{B}(\gamma_\mathrm{sol},-\mathbf{e}')
\begin{pmatrix}
c_s(\mathbf{\lambda}(t))t \\ \mathbf{x}
\end{pmatrix}.
\end{equation}

The initial SCO is Lorentz-type transformed in $-\mathbf{e}'$-direction. This does not translate into a movement of the SCO in $-\mathbf{e}'$-direction, but rather to an acceleration in said direction. This is derived in detail in this section.

Relating the SCO solution to the SCO at rest reveals 
\begin{equation}
\begin{pmatrix}
c_{s,0}t'' \\\mathbf{x}''
\end{pmatrix}
=
\mathbf{B}(\gamma_i,\mathbf{n})\mathbf{B}(\gamma_\mathrm{sol},-\mathbf{e}')
\begin{pmatrix}
c_s(\mathbf{\lambda}(t))t \\ \mathbf{x}
\end{pmatrix},
\end{equation}
with the Lorentz matrix $\mathbf{B}$ as defined in Eq.\ \ref{Lor}.

As explained in Sec.\ \ref{subsec:wigner}, these two matrices can be written as the combination of a single Lorentz transformation and a spatial rotation. The parameters of the single Lorentz transformation can be obtained by comparing the equation above with Eq.\ \ref{twolorentz}, resulting in the parameters described in Eq.\ \ref{WigSix} to Eq.\ \ref{WigSix2}. The combined $\gamma(t)$ is given by (use $\beta(t)$ accordingly):
\begin{equation}\label{combinedgamma}
\gamma(t) = \gamma_i\gamma_\mathrm{sol}(1-\beta_i\beta_\mathrm{sol}(\mathbf{n}\cdot\mathbf{e}')).
\end{equation}

The direction of motion of the SCO is given by the unit-vector $\mathbf{p}(t)$. For future reference, the unit-vector $\mathbf{o}(t)$ is given as well:
\begin{multline}\label{pando}
\mathbf{p}(t) = \frac{\beta_i\gamma_i\mathbf{n} + [\beta_i\gamma_i(\gamma_\mathrm{sol}-1)(\mathbf{n}\cdot\mathbf{e}') - \gamma_i\gamma_\mathrm{sol}\beta_\mathrm{sol}]\mathbf{e}'}{\beta(t)\gamma(t)}, 
\end{multline}

\begin{multline}
\mathbf{o}(t) =  \frac{-\beta_\mathrm{sol}\gamma_\mathrm{sol}\mathbf{e}' + [-\beta_\mathrm{sol}\gamma_\mathrm{sol}(\gamma_i-1)(\mathbf{e}'\cdot\mathbf{n}) + \gamma_\mathrm{sol}\gamma_i\beta_i]\mathbf{n}}{\beta(t)\gamma(t)}.\label{pando2}
\end{multline}

The unit-vector $\mathbf{p}(t)$, the direction of motion of the SCO solution, is equal to $\mathbf{n}$ at $t=0$. As time proceeds, it shifts towards $-\mathbf{e}'$. The unit-vector $\mathbf{o}(t)$ also starts as $\mathbf{n}$ at time $t=0$ and shifts towards $-\mathbf{e}'$, but at a different rate.

The trajectory of the SCO solution $\rho_\mathrm{sol}$ is completely described by the direction of motion $\mathbf{p}(t)$ and the velocity, which can be found in $\gamma(t)$ or $\beta(t)$. Similar to a velocity vector, it is possible to define the vector
$\vec{\beta}(t) = \beta(t)\mathbf{p}(t)$, called $\beta$-vector. In this case, the initial $\beta$-vector is $\vec{\beta}_i=\beta_i \mathbf{n}$. Starting with the unit-vector $\mathbf{p}(t)$ and $\gamma(t)$ given in Eq.\ \ref{combinedgamma}, the $\beta$-vector can be found to be 
\begin{equation}\label{finbeta}
\vec{\beta}(t) = \frac{\vec{\beta}_i + [(\gamma_\mathrm{sol}-1)
(\vec{\beta}_i\cdot\mathbf{e}') - \gamma_\mathrm{sol}\beta_\mathrm{sol}]\mathbf{e}'}{\gamma_\mathrm{sol}(1-\beta_\mathrm{sol}(\vec{\beta}_i\cdot\mathbf{e}'))}.
\end{equation}

Eq.\ \ref{finbeta} describes the motion of the SCO in a speed of propagation field with constant spatial gradient. The velocity is given by $|\vec{\beta}|(t)$ as a ratio with the local speed of propagation. The direction of motion is given by the direction of $\vec{\beta}(t)$. The advantage of Eq.\ \ref{finbeta} is that the local speed of propagation does not appear explicitly. Given the knowledge about the motion for all times, it is possible to calculate the momentary changes in $\vec{\beta}(t)$ by finding the differential equation for which Eq.\ \ref{finbeta} is a solution. This differential equation is
\begin{equation}\label{betadiff}
\dot{\vec{\beta}}(t) = -\mathbf{c}_s' + \vec{\beta}(t)[\mathbf{c}_s'\cdot\vec{\beta}(t)].
\end{equation}

This can be checked by using the relations in Eq.\ \ref{solrelations} to find the time derivative of Eq.\ \ref{finbeta} as
\begin{align}
\dot{\vec{\beta}}(t) ={}& c_s'\left[\frac{[\gamma_\mathrm{sol}\beta_\mathrm{sol}(\vec{\beta}_i\cdot\mathbf{e}') - \gamma_\mathrm{sol}]\mathbf{e}'}{\gamma_\mathrm{sol}(1-\beta_\mathrm{sol}(\vec{\beta}_i\cdot\mathbf{e}'))} \right.\nonumber\\
& \left.- \vec{\beta}\frac{\gamma_\mathrm{sol}\beta_\mathrm{sol} - \gamma_\mathrm{sol}(\vec{\beta}_i\cdot\mathbf{e}')}{\gamma_\mathrm{sol}(1-\beta_\mathrm{sol}(\vec{\beta}_i\cdot\mathbf{e}'))}\right],\nonumber\\
={}& c_s'[-\mathbf{e}' + \vec{\beta}(t)[\mathbf{e}'\cdot\vec{\beta}(t)]].
\end{align}

Eq.\ \ref{betadiff} describes the momentary changes of $\vec{\beta}(t)$ in a speed of propagation field with constant spatial gradient. It can be extended to a speed of propagation field with a changing spatial gradient. In this case, the spatial gradient $\mathbf{c}_s'$ is allowed to change in space as $\mathbf{c}_s'(\mathbf{\lambda}(t))$, 
\begin{equation}\label{betadiffcomp}
\dot{\vec{\beta}}(t) = -\mathbf{c}_s'(\mathbf{\lambda}(t)) + \vec{\beta}(t)[\mathbf{c}_s'(\mathbf{\lambda}(t))\cdot\vec{\beta}(t)].
\end{equation}

Eq.\ \ref{finbeta} depends on the vector $\vec{\beta}$ itself. The direction of motion with respect to the spatial gradient, as well as the ratio of the velocity with the local speed of propagation are important. This is an interesting feature. In fact, the second term of Eq.\ \ref{betadiff} ensures that the magnitude of $\vec{\beta}$ stays below unity. This can be seen by calculating the time-rate of change of the magnitude of $\vec{\beta}$ as
\begin{equation}\label{betadiffmag}
\dot{|\vec{\beta}|}\,(t) = \frac{\vec{\beta}(t)\cdot\dot{\vec{\beta}}(t)}{\beta(t)} = -(\mathbf{c}_s'\cdot\mathbf{p}(t))(1-\beta^2(t)).
\end{equation}

The change in magnitude $|\vec{\beta}|(t)$ approaches zero for $\beta(t)\to 1$. The SCO is not able to travel faster than the local speed of propagation. The SCO is not only described by its trajectory. It also undergoes spatial contraction, time dilation and spatial rotation due to Thomas-Wigner rotation. These effects are described later.

The differential equation for $\vec{\beta}(t)$ can also be written as an acceleration $\mathbf{a}(t)$; i.e. the change of the velocity vector with time as
\begin{equation}\label{accel}
\mathbf{a}(t) = c_s(\mathbf{\lambda}(t))\{-\mathbf{c}_s'(\mathbf{\lambda}(t)) + 2\vec{\beta}(t)[\mathbf{c}_s'(\mathbf{\lambda}(t))\cdot\vec{\beta}(t)]\},
\end{equation}
with the local value for the speed of propagation $c_s$.

\subsubsection{Dimensionless Equation of Motion}

With the same time substitution, $a=c_s't$, as performed in Sec.\ \ref{subsec:dimensionless}, the differential equation for $\vec{\beta}(t)$ in Eq.\ \ref{betadiff} can be described as a dimensionless equation. 
\begin{equation}\label{dimensionlesseqn}
\vec{\beta}_a(a) = -\mathbf{e}'(\mathbf{\lambda}(a)) + \vec{\beta}(a)[\mathbf{e}'(\mathbf{\lambda}(a))\cdot\vec{\beta}(a)].
\end{equation}

This is an interesting feature, as no constants of nature are needed. Only the properties of the medium at the position of the SCO are used. With no dependence on space and time, this equation does not depend on any definition thereof.

\subsubsection{SCOs travelling at the Local Speed of Propagation -- Photon-Prototype}
\label{sec:photon}

The motion of an SCO with a velocity close to the local speed of propagation can be examined by taking the limit $\beta\to 1$. Starting with the acceleration given in Eq.\ \ref{accel}, it simplifies to 
\begin{equation}\label{photonacc}
\mathbf{a}(t) = c_s(\mathbf{\lambda}(t))\{-\mathbf{c}_s'(\mathbf{\lambda}(t)) + 2\mathbf{s}(t)[\mathbf{c}_s'(\mathbf{\lambda}(t))\cdot\mathbf{s}(t)]\}.
\end{equation}

An SCO travelling at the local speed of propagation keeps travelling at the (changing) local speed of propagation. This can be seen in the change of the magnitude of the velocity,
\begin{equation}
\dot{v}(t) = \mathbf{a}(t)\cdot\mathbf{s}(t) = c_s(\mathbf{\lambda}(t))\mathbf{c}_s'(\mathbf{\lambda}(t))\cdot\mathbf{s}(t) = \frac{\mathrm{d}}{\mathrm{d}t}[c_s(\mathbf{\lambda}(t))].
\end{equation}

The magnitude of the SCO's velocity changes by the same amount as the speed of propagation field itself due to the motion of the SCO through space. The acceleration in Eq.\ \ref{photonacc} also contains information about the change in direction of the SCO travelling at the speed of propagation:
\begin{align}
\dot{\mathbf{s}}(t) ={}& \frac{\mathrm{d}}{\mathrm{d}t}\left[\frac{\mathbf{v}(t)}{v(t)}\right] = \frac{\mathbf{a}(t)}{v(t)} - \dot{v}(t)\mathbf{s}(t) \\
={}& -\mathbf{c}_s'(\mathbf{\lambda}(t)) + \mathbf{s}(t)[\mathbf{c}_s'(\mathbf{\lambda}(t))\cdot\mathbf{s}(t)],
\end{align}
which is equivalent to the change in direction of a plane wave constituent in an inhomogeneous speed of propagation field given in Eq.\ \ref{schange}. An SCO travelling close to the local speed of propagation has similar properties as a plane wave. In the limit $\beta\to 1$, no proper time elapses for the SCO. The Lorentz(-type) transformations ensure the spatial confinement of the SCO. An SCO travelling close to the local speed of propagation might be a good candidate for a photon-prototype.

\subsubsection{Thomas Precession}
\label{subsec:thomasprecession}

To derive the equation of motion of an SCO in an arbitrary inhomogeneous speed of propagation field, Thomas-Wigner rotation has to be used to combine a Lorentz transformation and a Lorentz-type transformation. This causes a spatial rotation of the SCO given by the rotation matrix $\mathbf{R}(\vec{\alpha})$ as described in Sec.\ \ref{subsec:wigner}. The vector $\vec{\alpha}$ is given in axis-angle-representation. The direction of $\vec{\alpha}$ is the rotation axis, the magnitude the rotation angle. 

The spatial rotation is present when an SCO in motion is accelerated. Similar to the derivation of the equation of motion, it is possible to calculate the momentary rotation velocity of an accelerated SCO. The momentary rotation is calculated for an SCO in motion at time $t=0$ with the onset of a second Lorentz-type transformation. The rotation angle is the angle between the unit-vectors $\mathbf{p}$ and $\mathbf{o}$ as given in Eq.\ \ref{rotationparams} and Eq.\ \ref{rotationparams2},
\begin{equation}
\sin(\alpha(t)) = |\mathbf{p}(t)\times\mathbf{o}(t)|.
\end{equation}

At time $t=0$, the motion of the SCO is given by $\mathbf{n}$ and $\beta_i$. The unit-vectors $\mathbf{p}(t=0)$ and $\mathbf{o}(t=0)$ are both equal to $\mathbf{n}$ with $\alpha=0$. With the onset of the second active continuous Lorentz-type transformation, both $\mathbf{p}$ and $\mathbf{o}$ change. After a time step $\mathrm{d}t$, these unit-vectors can be calculated by performing a Taylor expansion to first order on Eq.\ \ref{pando} and Eq.\ \ref{pando2},
\begin{align}\label{taylorunit}
\mathbf{p}(\mathrm{d}t) ={}& \mathbf{n} - \frac{\mathbf{c}_s'\mathrm{d}t}{\beta_i} + \frac{(\mathbf{c}_s'\cdot\mathbf{n})\mathbf{n}\mathrm{d}t}{\beta_i} + \mathcal{O}(\mathrm{d}t^2), \\
\mathbf{o}(\mathrm{d}t) ={}& \mathbf{n} - \frac{\mathbf{c}_s'\mathrm{d}t}{\beta_i\gamma_i} + \frac{(\mathbf{c}_s'\cdot\mathbf{n})\mathbf{n}\mathrm{d}t}{\beta_i\gamma_i} + \mathcal{O}(\mathrm{d}t^2).\label{taylorunit2}
\end{align}

Both unit-vectors change in a similar manner in $-\mathbf{e}'$-direction. The change of the unit-vector $\mathbf{p}$ is larger than the change of the unit-vector $\mathbf{o}$. This opens up an angle between them. After a time step $\mathrm{d}t$, this angle is equal to $\sin({\alpha}(\mathrm{d}t)) = \alpha + \mathcal{O}(\mathrm{d}t^2)$,
\begin{multline}
\alpha(\mathrm{d}t) = \left|\left(\mathbf{n} - \frac{\mathbf{c}_s'\mathrm{d}t}{\beta_i} + \frac{(\mathbf{c}_s'\cdot\mathbf{n})\mathbf{n}\mathrm{d}t}{\beta_i}\right)\right.\nonumber\\
\left.\times\left(\mathbf{n} - \frac{\mathbf{c}_s'\mathrm{d}t}{\beta_i\gamma_i} + \frac{(\mathbf{c}_s'\cdot\mathbf{n})\mathbf{n}\mathrm{d}t}{\beta_i\gamma_i}\right)\right| + \mathcal{O}(\mathrm{d}t^2).
\end{multline}

After some simplifications, this results in a differential equation for the instantaneous rotation of the accelerated SCO at time $t=0$,
\begin{equation}\label{thomas1}
\dot{\alpha}(t=0) = \frac{|\mathbf{c}_s'\times\mathbf{n}|}{\beta_i}\left(1-\frac{1}{\gamma_i}\right).
\end{equation}

At any time $t$ along the trajectory of an SCO, the current movement can be interpreted as an initial motion, with the SCO being affected by the onset of a Lorentz-type transformation. Thus, Eq.\ \ref{thomas1} can be extended to describe the instantaneous rotation of an SCO with $\vec{\beta}$ being affected by a spatial gradient $\mathbf{c}_s'$ as

\begin{equation}
\dot{\alpha}(t) = \frac{|\mathbf{c}_s'\times\vec{\beta}(t)|}{\beta^2(t)}\left(1-\frac{1}{\gamma(t)}\right).
\end{equation}

This rotation is called Thomas precession. It is usually written using velocity and acceleration. This can be done for an SCO as well, see below. First, using Eq.\ \ref{accel}, it can be seen that
\begin{equation}
\frac{\mathbf{a}(t)\times\mathbf{v}(t)}{c_s^2} = -\mathbf{c}_s'\times\vec{\beta}(t).
\end{equation}

The rotation axis is given by the cross product of $\mathbf{o}$ and $\mathbf{p}$. Using Eq.\ \ref{taylorunit} and Eq.\ \ref{taylorunit2}, it can be seen that the unit-vector $\mathbf{o}$ points more in the direction of the initial motion, while the unit-vector $\mathbf{p}$ points more in the direction of the acceleration. Thus, the rotation axis is given by
\begin{equation}
\frac{\mathbf{o}(t)\times\mathbf{p}(t)}{|\mathbf{o}(t)\times\mathbf{p}(t)|} = \frac{\mathbf{v}(t)\times\mathbf{a}(t)}{|\mathbf{v}(t)\times\mathbf{a}(t)|}.
\end{equation}

The instantaneous rotation can be expressed with a rotation vector $\vec{\omega}$, depending on the current velocity and acceleration,
\begin{equation}\label{resultthomas}
\vec{\omega}(t) =  \frac{\mathbf{v}(t)\times\mathbf{a}(t)}{v^2(t)}\left(1-\frac{1}{\gamma(t)}\right).
\end{equation}

This is the same expression as the Thomas precession found in special relativity \citep{precession}. It emerges naturally for an SCO in motion being accelerated. It represents a real rotation in space.

\subsubsection{Proper Time and Gravitational Redshift}
\label{subsec:inhomogeneouspropertime}

The trajectory of the accelerated SCO is derived in the previous subsections. To find other properties of an accelerated SCO, the results found in Sec.\ \ref{subsec:oscinmotion}, concerning the shape of the SCO and its oscillation, have to be modified to include the Lorentz-type transformation. 

The accelerated SCO is compared to a reference SCO at rest at a position in space with reference speed of propagation $c_\mathrm{ref}$. This can be any point in space. The reference SCO is used to define the coordinate time. The path $\mathbf{\lambda}(t)$ and the $\beta$-vector $\vec{\beta}(t)$ are considered to be solutions to the differential equation in Eq.\ \ref{betadiff}. 

From Eq.\ \ref{Wig}, considering the Lorentz-type transformation, by substituting $c_st\to c_s(\mathbf{\lambda}(t))t$, the active time transformation between the reference SCO and the accelerated SCO is
\begin{equation}
t'' = \frac{\gamma c_s(\mathbf{\lambda}(t))}{c_\mathrm{ref}}t - \frac{\gamma(\vec{\beta}(t)\cdot\mathbf{x})}{c_\mathrm{ref}}.
\end{equation}

This equation describes the linear relation between the old time coordinate and the new time coordinate at the position $\mathbf{x}$ in space for an accelerated SCO at position $\mathbf{\lambda}(t))$. Following the center of the accelerated SCO, $\mathbf{x}_c(t) = \vec{\beta}(t)c_s(\mathbf{\lambda}(t))t$, to map the center of the stationary SCO to the center of the accelerated SCO leads to the proper time of the accelerated SCO as
\begin{equation}\label{propcs0}
\tau = \frac{c_s(\mathbf{\lambda}(t))}{\gamma c_\mathrm{ref}}t.
\end{equation}

This equation can also be used to compare two SCOs at rest at different positions in space. Through some mechanism, an SCO might be moved to another position, being at rest at the beginning and the end. The SCOs at the beginning and at the end have different oscillation frequency if their respective local speed of propagation values are different. This can be interpreted as gravitational red- and blueshift of particles.

As mentioned above, the term $c_\mathrm{ref}$ in  Eq.\ \ref{propcs0} depends on the definition of coordinate time, which is better represented by Eq.\ \ref{propcs0}. There is no special speed of propagation value. There is only a reference speed of propagation which is connected to the definition of coordinate time.

More similarities to general relativity can be found. This concerns the geodesic equation describing the trajectory and the proper time of particles in general relativity. It is possible to describe the change in the relation between proper time and coordinate time on the trajectory of the SCO. This can be written as the second derivative of the coordinate time w.r.t.\ the proper time of the SCO. For an SCO in an inhomogeneous speed of propagation field, the relation between proper time and coordinate time is given by Eq.\ \ref{propcs0}. For differential time steps, it can be written as
\begin{equation}
\frac{\mathrm{d}t}{\mathrm{d}\tau} = \frac{c_\mathrm{ref}}{c_s(\mathbf{\lambda}(t))\sqrt{1-\beta^2(t)}}.
\end{equation}

A second derivate of the coordinate time $t$ w.r.t.\ the proper time $\tau$ is found via the following relation:
\begin{equation}\label{proptimederi}
\frac{\mathrm{d}^2t}{\mathrm{d}\tau^2} = \frac{1}{2}\frac{\mathrm{d}}{\mathrm{d}t}\left(\frac{\mathrm{d}t}{\mathrm{d}\tau}\right)^2 = -\frac{2\gamma^2(t)c^2_\mathrm{ref}}{c_s^2(\mathbf{\lambda}(t))}[\mathbf{c}_s'(\mathbf{\lambda}(t))\cdot\vec{\beta}(t)],
\end{equation}
using the differential equation for the change in $\vec{\beta}(t)$ given in Eq.\ \ref{betadiffmag}. This function describes how the relation between proper time and coordinate time changes along the trajectory. There are two effects; a change in the local value of the speed of propagation field and a change in $\vec{\beta}$ due to the spatial gradient in the speed of propagation field. They contribute equally to the change of $\mathrm{d}t/\mathrm{d}\tau$.

A second proper time derivative of the coordinate time as found in Eq.\ \ref{proptimederi} for an SCO is also found in the 0th component of the geodesic equation from general relativity. The next subsection compares the properties of SCOs and particles in general relativity in detail.

\subsubsection{Geodesic Trajectories}
\label{subsec:geodesictrajectory}
Starting with the acceleration of an SCO given in Eq.\ \ref{accel}, the time derivatives can be substituted by derivatives by the proper time $\tau$ using Eq.\ \ref{propcs0} and Eq.\ \ref{proptimederi}.
\begin{equation}\label{spatialpa}
\frac{\mathrm{d}^2\mathbf{\lambda}(t)}{\mathrm{d}\tau^2} + c_s(\mathbf{\lambda}(t)) 
c_s'(\mathbf{\lambda}(t))\left(\frac{\mathrm{d}t}{\mathrm{d}\tau}\right)^2 = 0.
\end{equation}
Together with Eq.\ \ref{proptimederi} it can be shown that these equation correspond to the 0th and spatial components of the geodesic equation given by the spacetime defined by \begin{equation}
    g = \begin{pmatrix}
    c_s^2(\mathbf{x}) & 0 & 0 & 0 \\
    0 & -1 & 0 & 0 \\
    0 & 0 & -1 & 0 \\
    0 & 0 & 0 & -1
    \end{pmatrix}.
\end{equation}

This has now established that an SCO can exhibit the same properties as a particle in general relativity, including the same trajectory, the same concept of proper time and gravitational red- and blueshift. For an SCO, these properties emerge in euclidean space from simple wave mechanics.

\subsubsection{Remarks on Time-Dependence}

The speed of propagation field might also depend on time. In that case, the local parameters of the speed of propagation field $c_{s,0}$ and $\mathbf{c}_s'$ can be considered to be time-dependent. An SCO at a fixed time is subject to the local parameters of the speed of propagation field at that time. This has to be studied in detail, starting from the effect of a time-dependent speed of propagation field on the refraction of plane waves.

\subsubsection{Order of Magnitude of Spatial Gradient}

For SCOs with small velocity, the second term in the acceleration in Eq.\ \ref{accel} can be dropped. The order of magnitude of the spatial gradient of an inhomogeneous speed of propagation field can be approximated for different physical accelerations via $|\mathbf{a}(t)| = c_s(\mathbf{\lambda}(t))c_s'(\mathbf{\lambda}(t))$.

For gravity on Earth, it is assumed that the speed of propagation is approximately $c_{s,0}=3\times 10^8\,\frac{\mathrm{m}}{\mathrm{s}}$. The order of magnitude of the spatial gradient is
\begin{equation}
c_s' = \frac{g}{c_{s,0}} \approx 3 \times 10^{-8}\frac{\mathrm{m}/\mathrm{s}}{\mathrm{m}}.
\end{equation}

On Earth, the speed of propagation field would change in its 17th significant digit on a meter scale.

In the Milky Way, for gravitational accelerations on galaxy scales, the order of magnitude of the spatial gradient at the position of the Sun at $8\,\mathrm{kpc}$ is
\begin{equation}
c_s' \approx 10^{-18}\frac{\mathrm{m}/\mathrm{s}}{\mathrm{m}}.
\end{equation}

The speed of propagation field would change in its 27th significant digit on a meter scale. On kpc-scales, the speed of propagation field changes in the order of $10\,\mathrm{m}/\mathrm{s}$. These concepts might also be interesting to examine in the context of cosmological models.

\subsubsection{Remarks on Inhomogeneous Mediums}

The introduction of an inhomogeneous speed of propagation field leads to a particle-field interaction between SCO and medium. The SCO is accelerated, with neither applying the concepts of force nor mass. The acceleration naturally arises from basic principles, namely the shift in the region of constructive interference of the plane wave constituents of the SCO. Applying the theory presented in this contribution to reality, gravitational acceleration might also be interpreted in terms of a speed of propagation field  causing an acceleration of SCOs. However, the primary goal of this contribution is to provide an additional vantage point for viewing gravitation without contradicting the exiting interpretations. The results found with this manuscript may be useful for considering Einstein's field equation and Verlinde's holographic principle as potentially related with the wave-nature of matter. The results obtained may in principle be open to testing in an appropriately constructed classical laboratory.


\section{Conclusion}
\label{sec:conclusion}

Entities carrying energy, even when at rest, are constructed by superposition of plane waves. These SCOs obey the same special-relativistic rules as real matter particle do and are, by their wave nature, quantum mechanical in character. Introducing gradients in the speed of propagation, the continuous Ibn-Sahl--Snell law is derived. A complete description of the dynamics of SCOs in inhomogeneous speed of propagation fields is derived. The trajectory of an SCO is described by the differential equation for the $\beta$-vector in Eq.\ \ref{betadiff} or the acceleration in Eq.\ \ref{accel}. Neither of these equations requires the concepts of forces or masses. A dimensionless equation of motion can be found, see Eq.\ \ref{dimensionlesseqn}. This equation of motion is completely described by the local parameters of the speed of propagation field without the need of constants of nature. Only basic principles such as the refraction of plane waves in an inhomogeneous medium are employed. Thus, an SCO localised in a medium with a gradient in the speed of propagation will accelerate and obey an equation of motion as known for real matter particles in a gravitational fiel. 

SCOs exhibit the concept of proper time by defining the oscillation of their center as an internal clock. 
In this contribution it is shown that proper time is affected by the local speed of propagation value and the velocity of the SCO through the inhomogeneous medium, resulting in properties equivalent to general relativity, see Sec.\ \ref{subsec:inhomogeneouspropertime}.

Interestingly, calculating the geodesic equation for the corresponding metric in general relativity results in the same equation of motion as the one for an SCO derived by refracting plane waves in an inhomogeneous medium. In the case of flat-space metrics, the geodesic equation is equivalent to the mechanism of superimposing plane waves and describing the movement of the region of their constructive interference, see Sec.\ \ref{subsec:geodesictrajectory}. Whether or not this can be extended to curved-space metrics has to be studied in detail.

From direct calculations, other general relativistic properties are found for SCOs. These include the concept of gravitational red- and blueshift of particle-prototypes (Sec.~\ref{subsec:inhomogeneouspropertime}) and photon-prototypes (Sec.\ \ref{sec:photon}). Thomas precession from particle physics is also a result from direct calculations, see Sec.\ \ref{subsec:thomasprecession} and Eq.\ \ref{resultthomas}. It emerges naturally for an SCO in motion being accelerated and represents a real rotation in space. For an SCO these properties emerge from simple wave mechanics. These results might thus allow to observe general relativistic behaviour in classical wave experiments.

This contribution represents the first step to find a general Lorentz ether theory
(Sec.~\ref{subsec:history}), which includes the dynamics of SCOs, in
analogy to the role of general relativity in relation to special
relativity. The results of this
work may be relevant for interpreting gravitation as an emergent
property from the wave nature of matter but are also of interest in their own right. This work suggests that general relativistic effects might be accessible to classical wave experiments. Future research will involve the study of the mutual refraction of two SCOs which will necessitate the introduction of non-linear wave mechanics. It will be interesting to investigate if it is possible to place two SCOs into orbit about each other as a consequence of the mutual refraction of their constituent waves. If so, then this may constrain the equation of state of the medium in which the waves propagate which will be useful for constructing laboratory experiments. Intriguingly, it will be interesting to investigate if the non-linear wave mechanics in classical media might lead to deviations from the classical equations of motion of two SCOs in mutual orbit which may or may not resemble a missing mass problem in the hypothetical idealised toy universe used as a laboratory.


	
	\bibliography{refs}
	
\end{document}